\documentclass[twocolumn,aps, prb, showpacs, preprintnumbers, amsmath, amssymb, a4paper, superscriptaddress]{revtex4-1}

\usepackage{graphicx}
\usepackage{dcolumn}
\usepackage{bm}
\usepackage[percent]{overpic}
\usepackage{multirow}
\usepackage{color}

\begin{document}
\newcommand{\Arg}[1]{\mbox{Arg}\left[#1\right]}
\newcommand{\bb}{\mathbf}
\newcommand{\braopket}[3]{\left \langle #1\right| \hat #2 \left|#3 \right \rangle}
\newcommand{\braket}[2]{\langle #1|#2\rangle}
\newcommand{\be}{\[}
\newcommand{\br}{\vspace{4mm}}
\newcommand{\bra}[1]{\langle #1|}
\newcommand{\braketbraket}[4]{\langle #1|#2\rangle\langle #3|#4\rangle}
\newcommand{\braop}[2]{\langle #1| \hat #2}
\newcommand{\dd}[1]{ \! \! \!  \mbox{d}#1\ }
\newcommand{\DD}[2]{\frac{\! \! \! \mbox d}{\mbox d #1}#2}
\renewcommand{\det}[1]{\mbox{det}\left(#1\right)}
\newcommand{\ee}{\]} 
\newcommand{\eg}{\textbf{\\  Example: \ \ \ }}
\newcommand{\Imag}[1]{\mbox{Im}\left(#1\right)}
\newcommand{\ket}[1]{|#1\rangle}
\newcommand{\ketbra}[2]{|#1\rangle \langle #2|}
\newcommand{\kp}{\arccos(\frac{\omega - \epsilon}{2t})}
\newcommand{\ldos}{\mbox{L.D.O.S.}}
\renewcommand{\log}[1]{\mbox{log}\left(#1\right)}
\newcommand{\Log}{\mbox{log}}
\newcommand{\Modsq}[1]{\left| #1\right|^2}
\newcommand{\nb}{\textbf{Note: \ \ \ }}
\newcommand{\op}[1]{\hat {#1}}
\newcommand{\opket}[2]{\hat #1 | #2 \rangle}
\newcommand{\occ}{\mbox{Occ. Num.}}
\newcommand{\Real}[1]{\mbox{Re}\left(#1\right)}
\newcommand{\so}{\Rightarrow}
\newcommand{\sol}{\textbf{Solution: \ \ \ }}
\newcommand{\thetafn}[1]{\  \! \theta \left(#1\right)}
\newcommand{\tin}{\int_{-\infty}^{+\infty}\! \! \!\!\!\!\!}
\newcommand{\Tr}[1]{\mbox{Tr}\left(#1\right)}
\newcommand{\kb}{k_B}
\newcommand{\rad}{\mbox{ rad}}
\preprint{APS/123-QED}

\title{Electronic transport in disordered graphene antidot lattice devices}

\
\author{Stephen R. Power}
\email{spow@nanotech.dtu.dk}
\author{Antti-Pekka Jauho}
\affiliation{Center for Nanostructured Graphene (CNG), DTU Nanotech, Department of Micro- and Nanotechnology,
Technical University of Denmark, DK-2800 Kongens Lyngby, Denmark}

\date{\today}

\begin{abstract}
Nanostructuring of graphene is in part motivated by the requirement to open a gap in the electronic band structure.
In particular, a periodically perforated graphene sheet in the form of an antidot lattice may have such a gap.
Such systems have been investigated with a view towards application in transistor or waveguiding devices.
The desired properties have been predicted for atomically precise systems, but fabrication methods will introduce significant levels of disorder in the shape, position and edge configurations of individual antidots.
We calculate the electronic transport properties of a wide range of finite graphene antidot devices to determine the effect of such disorders on their performance.
Modest geometric disorder is seen to have a detrimental effect on devices containing small, tightly packed antidots which have optimal performance in pristine lattices.
Larger antidots display a range of effects which strongly depend on their edge geometry.
Antidot systems with armchair edges are seen to have a far more robust transport gap than those composed from zigzag or mixed edge antidots.
The role of disorder in waveguide geometries is slightly different and can enhance performance by extending the energy range over which waveguiding behaviour is observed.
\end{abstract}

                 
\maketitle

\section{Introduction}
\label{intro}

Much recent effort in graphene research has focused on attempts to introduce a bandgap into the otherwise semi-metallic electronic band structure of graphene. 
Such a feature would allow the integration of graphene, with its many superlative physical, electronic, thermal and optical properties, into a wide range of device applications.  \cite{riseofgraphene, neto:graphrmp}
In particular, the presence of a bandgap is a vital step in the development of a graphene transistor capable of competing with standard semiconductor-based devices.

Initial investigations into gapped graphene were primarily based around graphene nanoribbons, with the electron confinement induced by the presence of crystalline edges predicted to introduce a bandgap similar to that found in many carbon nanotubes.\cite{Nakada:1996ribbons, ezawa:ribbonwidth, brey_electronic_2006, Son:ribbonenergygaps} 
More recent efforts have turned towards graphene superlattices, where the imposition of a periodic perturbation of the graphene sheet is also predicted to open up a bandgap. 
The periodic perforation of a graphene sheet, to form a so-called graphene antidot lattice (GAL) or nanomesh, is one such implementation of the latter technique. \cite{Pedersen:GALscaling, Pedersen2008, Furst2009, Furst2009PRB, Vanevic2009, Rosales2009, Zheng2009, Ritter2009, ClarSextetGAL, Ouyang2011, Tue:finiteGAL, Pedersen2012a, GALWG, Chang2012, Yuan_GAL_disorder, Yuan_GAL_screening, Liu2013, HungNguyen2013, Trolle-spinGAL, Dvorak2013, Ji2013, Brun2014, Bieri2009, Kim2010, Bai2010, Kim2012, Shen2008, Eroms2009, Begliarbekov2011, Giesbers2012, Oberhuber2013, Xu2013}
Periodic gating\cite{Low2011, Pedersen2012} and strain\cite{pereira_strain_2009, pereira_tight-binding_2009} are other possible routes that have been suggested.

Theoretical studies of GAL-based systems have suggested that the bandgap behaviour in many cases follows a simple scaling law relating the period of the perturbation and the antidot size.\cite{Pedersen:GALscaling} 
In other cases, more complex symmetry arguments involving the lattice geometry\cite{ClarSextetGAL, Ouyang2011, Liu2013} or the effect of edge states\cite{Vanevic2009, Tue:finiteGAL, Trolle-spinGAL} can be employed to predict the presence and magnitude of the bandgap. 
Furthermore, it is predicted that only a small number of antidot rows are required to induce bulk-like transport gaps\cite{Tue:finiteGAL, Pedersen2012a, HungNguyen2013}, suggesting the use of GALs in finite barrier systems which do not suffer from the Klein-tunnelling driven barrier leakage expected for gated systems.\cite{katsnelson_chiral_2006}
Indeed, the potential barrier efficacy of GALs suggests application in electron wave-guiding\cite{GALWG}, in analogy with photonic crystals where antidot lattice geometries have also been considered.\cite{Chutinan2000}

However, many of these potential device applications are predicated on atomically precise graphene antidot structures, whereas experimental fabrication (primarily involving block copolymer etching\cite{Bieri2009, Kim2010, Bai2010, Kim2012} or electron beam lithography\cite{Shen2008, Eroms2009, Begliarbekov2011, Giesbers2012, Oberhuber2013, Xu2013} techniques) will inevitably introduce a degree of imperfection and disorder into the system. 
It may however be possible to control the antidot edge geometries to some extent by heat treatment\cite{Jia2009, Xu2013} or selective etching\cite{Oberhuber2013, Pizzocchero2014}. 
Much like the properties of nanoribbons were found to be greatly affected by disorder,\cite{mucciolo:graphenetransportgaps, evaldsson:ribbonedgeanderson} recent studies suggest that the electronic and optical properties of GALs may also be strongly perturbed.\cite{Yuan_GAL_disorder, Yuan_GAL_screening, HungNguyen2013, Ji2013}
We should therefore expect that the transport properties and device fidelity of the systems described above will depend on the degree of disorder present in the antidot lattice.

Motivated by this concern we have simulated a wide range of finite GAL devices, in both simple-barrier and waveguide geometries, with various disorder types and strengths. 
We find that the geometries predicted to give the largest bandgaps, namely those with a dense array of small holes, are particularly susceptible to the effects of disorder and that transport gaps are quickly quenched as leakage channels form at energies in the bandgap. 
Geometric disorder, consisting of fluctuations in the positions and sizes of the antidots, is found to have a particularly dramatic effect. 
However, for larger antidots the signatures of such disorders are found to be strongly dependent on the edge geometry of the individual antidots.
We observe different behaviour when the antidot edge atoms have armchair or zigzag configurations, or alternating sequences of both. 
The remainder of the paper is organised as follows: Section \ref{section_model} introduces the geometry of the systems under consideration, the details of the electronic model and transport calculations and the types of disorder that will be applied. 
Section \ref{sec_barrier_results} examines in detail the transport properties of finite-length barrier geometry devices, starting with pristine systems (\ref{subs_sec_clean}). 
Small antidot systems with edge roughness (\ref{subsec_edge_roughness}) and geometric disorders (\ref{subsec_geo_dis}) are presented, before we consider larger antidot systems and edge geometry effects with and without the presence of disorder (\ref{subsec_edge_effects}). 
Section \ref{sec_WG} considers devices with waveguide geometries and with the various types of antidots and disorders discussed previously for barrier devices. 
Finally, Section \ref{sec_conclude} summarizes and discusses the implications of our results in the context of device optimization and recent experimental progress.

\section{Model}
\label{section_model}

\begin{figure}
\centering
\begin{tabular}{c c c}
  \multicolumn{2}{c}{a)\,\includegraphics[width =0.28\textwidth]{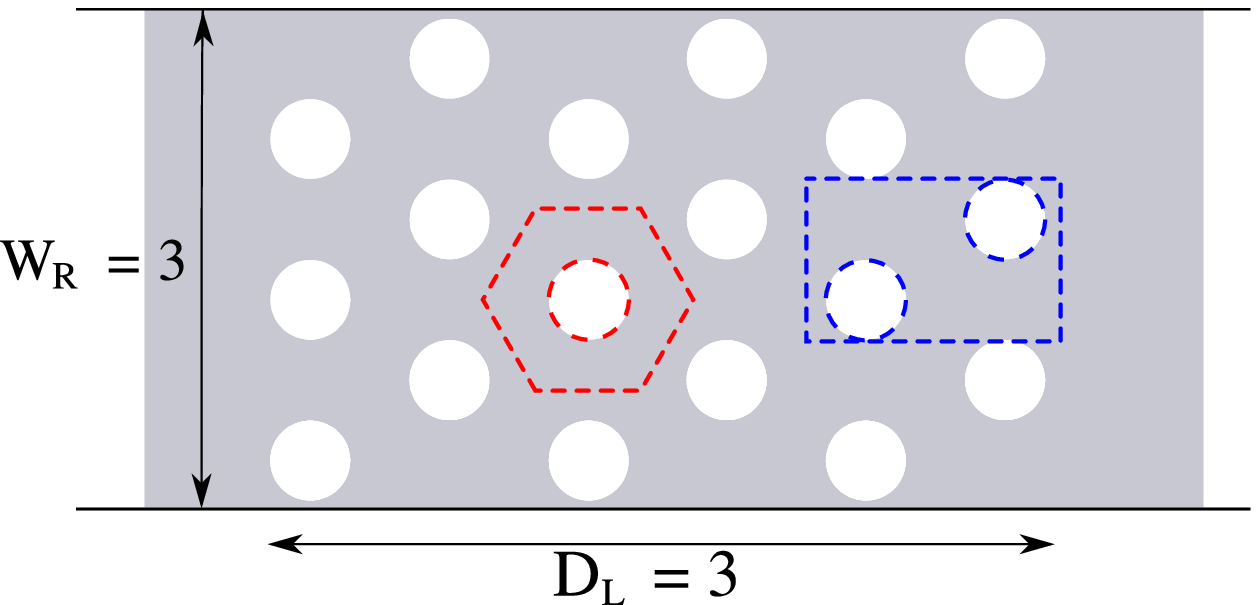}}&
 b) \includegraphics[width =0.155\textwidth]{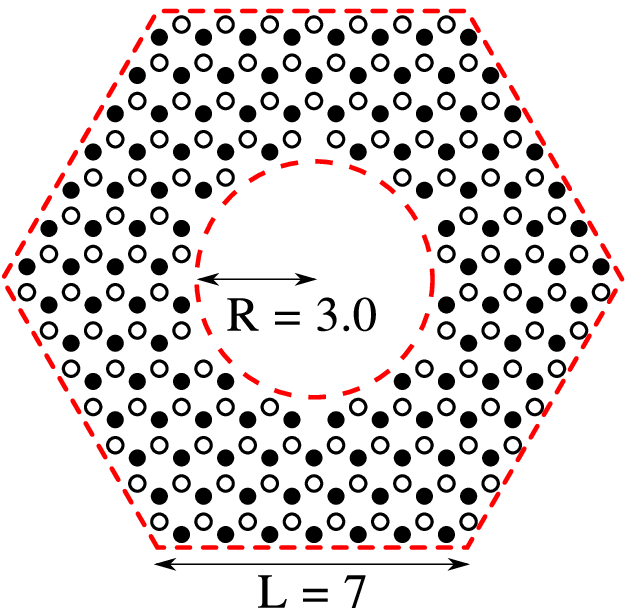} \\
  & & \\
\includegraphics[width =0.125\textwidth]{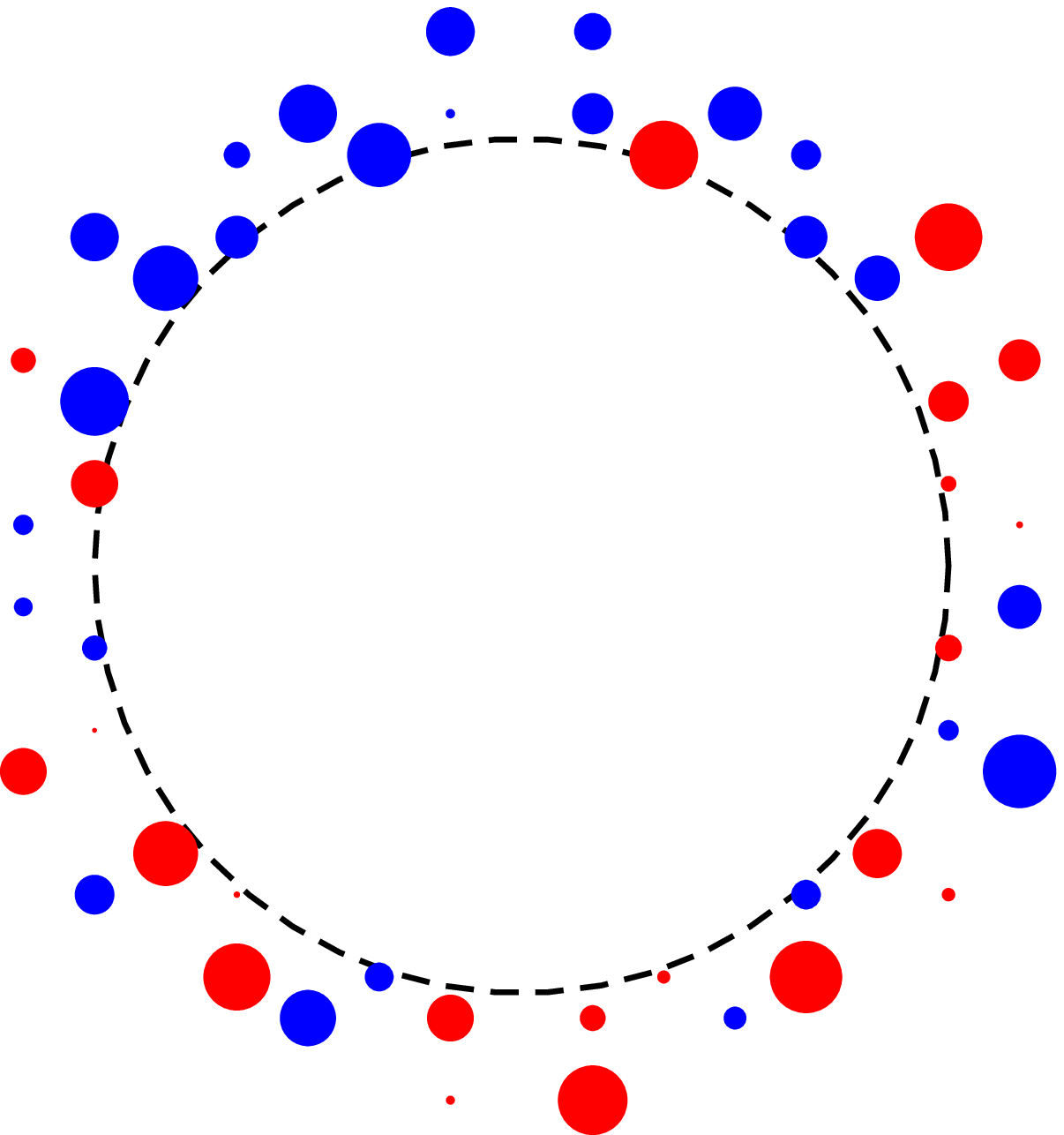} &
\,\includegraphics[width =0.145\textwidth]{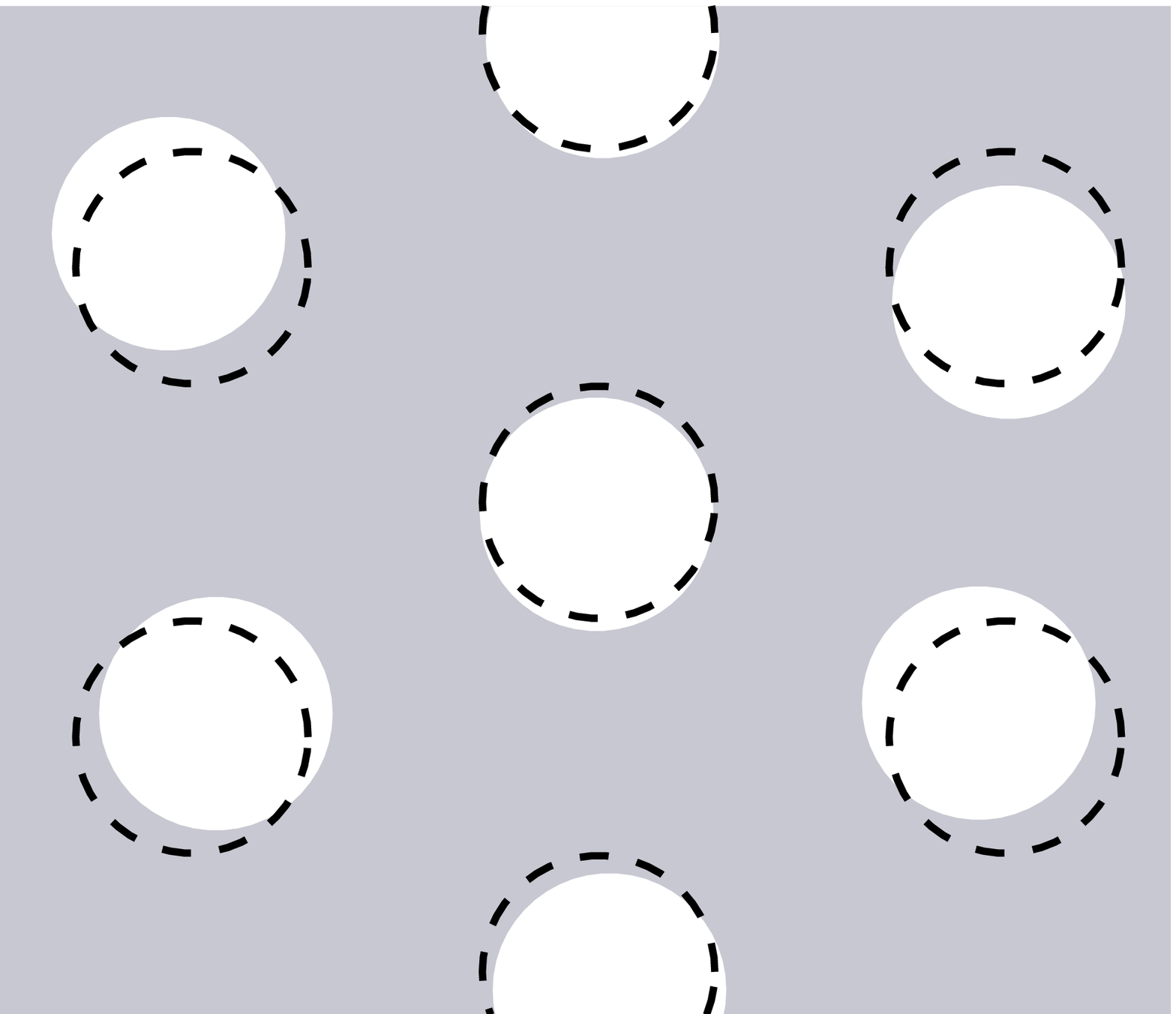} & 
\,\includegraphics[width =0.145\textwidth]{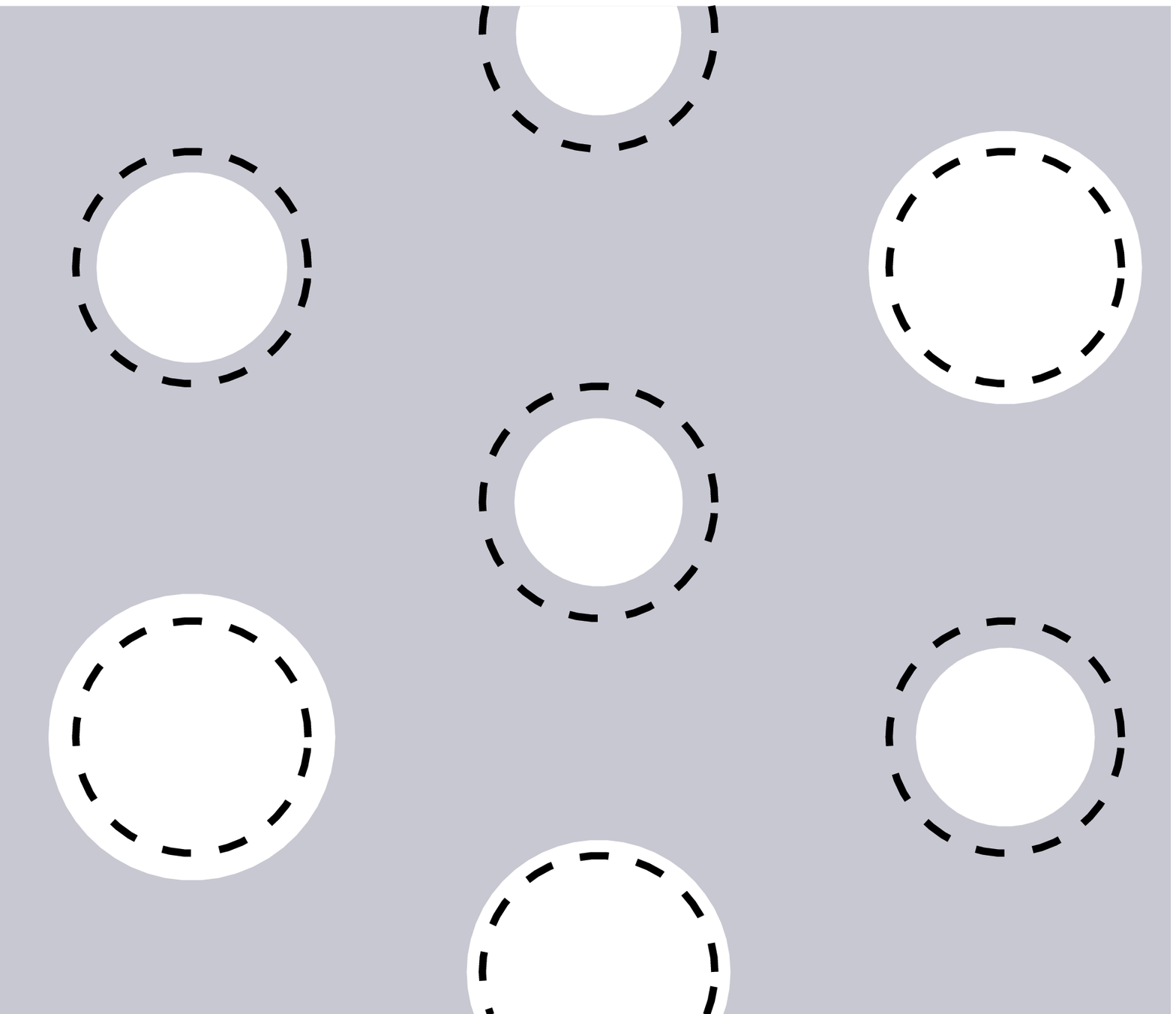} \\ 
c) Edge roughness & d) Position & e) Radial \\
\end{tabular}
\caption{a) Schematic of a simple antidot barrier device in a graphene nanoribbon. The white circles represent regions where carbon atoms have been removed from the lattice. The red and blue dashed lines show the single-antidot and double-antidot unit cells respectively. b) The atomic structure for a single circular $\{7, 3.0\}_C$ antidot. Schematics of c) antidot edge roughness, d) position and e) radial disorders. The dashed lines in each case represent the edge of the pristine antidot. The blue (red) circles in panel c) represent carbon atoms with an added positive (negative) onsite potential term proportional to the circle size.}
\label{fig_schematic}
\end{figure}

\subsection{Antidot and device geometry}
\label{model_shapes}

The systems we consider in this work consist of finite-length device regions connected to semi-infinite leads. 
A schematic geometry of a GAL barrier system is presented in Fig. \ref{fig_schematic}a). 
The device region is built from rectangular cells (blue-dashed lines) containing two possible antidot sites from a triangular GAL. 
This lattice geometry is selected as it is predicted to always open an electronic bandgap, unlike square and rotated triangular lattices where particular unit cell sizes may give rise to gapless systems\cite{ClarSextetGAL}.
The rectangular cell is exactly twice the size of the hexagonal unit cell of such a lattice which is shown by red dashed lines.
The \emph{device length} $D_L$ and \emph{ribbon width} $W_R$ shown in Fig. \ref{fig_schematic}a) are given in units of the rectangular cell's length and width respectively.
The antidot lattices considered are characterised by the size, shape and spacing of the antidots. 
We use the notation $\{L, R\}_x$, where $L$ and $R$ are the side length of the unit cell and antidot radius respectively, both given in units of the graphene lattice parameter $a=2.46\mathrm{\AA}$. 
$x = C, Z, A$ specifies whether the antidot has a circular shape ($C$) or a hexagonal shape with zigzag ($Z$) or armchair ($A$) geometry. 
The electronic bandgap of a triangular lattice of such antidots is predicted\cite{Pedersen:GALscaling} to scale as $E_G \sim \frac{R}{L^2}$.
Fig. \ref{fig_schematic}b) shows the atomic structure of a unit cell of the $\{7, 3.0\}_C$ antidot lattice which will be considered throughout much of this work. 
The geometry of hexagonal antidots will be discussed later in the paper. 
The rectangular unit cells composing the device region have length $3La$ and width $\sqrt{3} La$ and contain $12 L^2$ carbon atoms in the absence of an antidot ($R=0$). 
$L$ is always an integer, but $R$ can take non-integer values. 
The semi-infinite leads in the system are wide zigzag graphene nanoribbons (ZGNRs) and in the $R=0$ case the entire device can be considered as a pristine $2 L\, W_R\,$-ZGNR, using the usual nomenclature where the integer counts the number of zigzag chains across the ribbon width.

\subsection{Electronic and transport calculations}
The electronic structure of the graphene systems investigated is described by a single $\pi$-orbital nearest-neighbour tight-binding Hamiltonian
\begin{equation}
 H = \sum_{<ij>} t_{ij} \, {\hat c}_{i}^\dag \, {\hat c}_{j} \;\,,
\label{hamiltonian}
\end{equation}
where the sum is taken over nearest-neighbour sites only.
The only non-zero element, in the absence of disorder, is $t = -2.7\mathrm{eV}$ for nearest-neighbour sites.
Throughout this work we will use $|t|$ as the unit of energy.
Carbon atoms can be removed from the antidot regions by removing the associated rows and columns from the system Hamiltonian.
Carbon atoms with only a single remaining neighbouring atom are also removed from the system.
Any dangling sigma bonds for a carbon atom with only two neighbouring carbon atoms are assumed to be passivated with Hydrogen atoms so that the $\pi$ bands are unaffected.

Transport quantities are calculated using recursive Green's function techniques. 
The device region is decomposed into a series of chains which are connected from left to right to calculate the Green's functions required.
The semi-infinite leads are constructed using an efficient decimation procedure which takes advantage of system periodicity.\cite{Sancho-Rubio}
A general overview of these techniques applied to graphene systems is given in Ref. [\onlinecite{Lewenkopf2013}].
The zero-temperature conductance is given by the Landauer formula\cite{DattaBook} $G = \frac{2e^2}{h} T$ , where the transmission coefficient is calculated using the Caroli formula
\begin{equation}
T(E) = \mathrm{Tr}\big[ \mathbf{G}^r(E) \mathbf{\Gamma}_R(E)  \mathbf{G}^a(E)  \mathbf{\Gamma}_L(E)\big], 
\end{equation}
where $ \mathbf{\Gamma}_{i}(E)$ ($i=L,R$) are the level width matrices describing the coupling of the device region to the left and right leads and $ \mathbf{G}^{r/a}$ is the retarded/advanced Green's function of the device region connected to the leads. 
The Green's functions required for this calculation are acquired from a single recursive sweep through the device.
When studying disordered systems it is usually necessary to take a configurational average over many instances of a particular set of disorder parameters in order to discern the overall trend.

In studying the antidot devices in this work, we will also examine local electronic properties in order to understand how transport through a device is affected by antidot geometry and disorder.
In particular we will map the \emph{local density of states} at a site $i$
\begin{equation}
\rho_{i} (E_F)= - \frac{1}{\pi} \mathrm{Im} \big[G_{ii} (E_F)\big]
\end{equation}
and the \emph{bond current} flowing between sites $i$ and $j$ 
\begin{equation}
J_{ij} (E_F)= - \frac{1}{\hbar} H_{ij} \mathrm{Im} \big[ \mathbf{G}^r \mathbf{\Gamma}_L \mathbf{G}^a \big]_{ij} \,,
\end{equation}
where $H_{ij}$ is the relevant matrix element of the Hamiltonian in Eq. \eqref{hamiltonian}.
In larger systems some spatial averaging of these quantities will be performed to ease visualisation.
The Green's functions required for these calculations are more computationally expensive than those for the transmission calculation, and involve sweeping forward and back through the device region using recursive methods.\cite{Lewenkopf2013}

\subsection{Modelling disorder}
\label{method_disorder}
In this work we will consider three types of disorder that are very difficult to avoid during the experimental fabrication of GAL devices.
Every circular antidot $i$ in a device is characterised by three parameters $(x_i, y_i, r_i)$ which represent the $x$ and $y$ coordinates of the antidot centre, and the antidot radius respectively. 
Hexagonal antidots are parameterised slightly differently and are discussed later.

The first type of disorder we consider is \emph{edge roughness}. 
This type of disorder places Anderson-like random potentials on sites within a small additional radius of the antidot edge and mimics, for example, minor atomic realignment, random edge passivations\cite{Ji2013}, vacancies\cite{HungNguyen2013},  or randomly adsorbed atoms near an antidot edge. 
It is characterised by two parameters, an additional radius $\delta R_{dis}$ and a strength $S$. 
All remaining carbon sites in the ring between $r_i$ and $r_i+\delta R_{dis}$ around an antidot centre are given a random onsite potential in the range $[-S, S]$, as illustrated in Fig. \ref {fig_schematic}c).

We also consider more serious geometric disorder in the form of fluctuations of the antidot position and radius.\cite{Yuan_GAL_disorder, Yuan_GAL_screening}
Antidot \emph{position disorder} is characterised by $\delta_{xy}$ such that the antidot site coordinates are chosen from the ranges 
\begin{align*}
 x_i^0 - \delta_{xy} & < x_i < x_i^0 + \delta_{xy} \\
 y_i^0 - \delta_{xy} & < y_i < y_i^0 + \delta_{xy} \,
\end{align*}
where $(x_i^0, y_i^0)$ are the coordinates of antidot $i$ in a pristine GAL system. 
Antidot \emph{radial disorder} allows for a similar fluctuation in the radius of the antidot so that the radius of each antidot is in the range
\begin{equation*}
 R - \delta_{r}  < r_i < R + \delta_{r}.
\end{equation*}

These theoretical disorders represent the deviations from atomically precise systems that are most likely to occur during experimental fabrication of antidots - namely the difficulties in keeping antidot edges perfectly clean and in controlling the position and size of each antidot with atomic precision.
We do not account for disorder away from the antidotted regions and assume that the graphene sheet into which the antidots are patterned is otherwise defect free. 
We thus neglect, for example, defects or adsorbants that may occur in rest of the graphene sheet due to the various etching procedures.
The electronic and optical properties of graphene antidots with disorder in the graphene regions have been considered in other works\cite{Yuan_GAL_disorder, Yuan_GAL_screening}, but here we focus entirely on the effects of antidot disorder.

\section{Finite length GAL barriers}
\label{sec_barrier_results}

\subsection{Clean barriers}
\label{subs_sec_clean}

\begin{figure}
\centering
\includegraphics[width =0.45\textwidth]{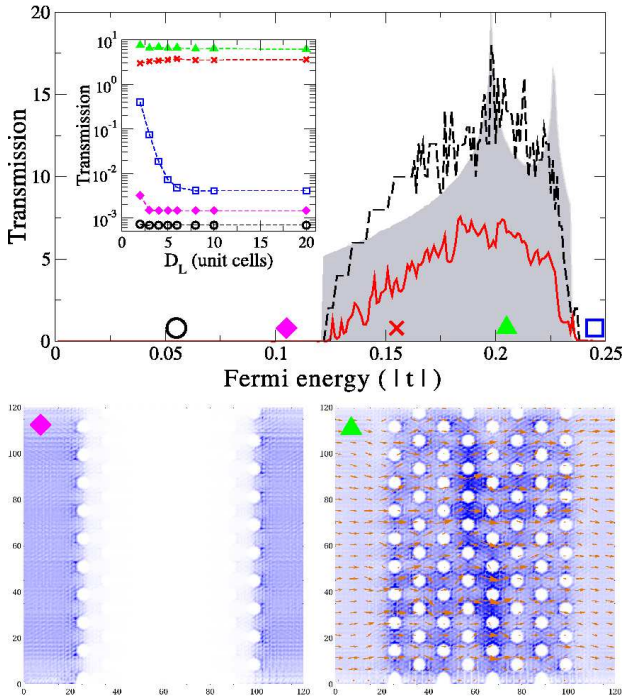}
\caption{Top panel shows the transmission through infinite and finite $\{7,3.0\}_C$ GAL barriers. 
The black dashed curve shows the transmission for the infinite GAL ribbon of width $W_R=10$. 
The solid red curve show the transmission through a system with clean graphene leads and a device length $D_L = 4$. 
The grey shaded area shows the (arbitarily scaled) DOS of the corresponding infinite GAL sheet. 
The inset shows the transmission as a function of $D_L$ calculated at different energies shown by the corresponding symbols on the energy axis of the main plot. 
The transmissions in the inset have been averaged over a narrow energy window as discussed in the main text.
The bottom panels show LDOS maps (blue shading) and current distributions (orange arrows) through the finite device at $E_{F}=0.1|t|$ (left) in the GAL bandgap and $E_{F}=0.2|t|$ (right) in the conducting range of the GAL.}
\label{fig_barrier_inf}
\end{figure}

We begin by examining systems built from the $\{7,3.0\}_C$ antidot lattice shown in Fig. \ref {fig_schematic}c). 
Lattices composed of such small, tightly-packed antidots have been the focus of recent theoretical investigation due to the large electronic bandgaps which follow due to the scaling law discussed earlier.\cite{Pedersen:GALscaling} 
A recent study suggests that only very few rows of antidots are required to achieve a transport gap in a finite GAL barrier which is identical to the bandgap of the corresponding infinite GAL system\cite{Tue:finiteGAL}. 
This behaviour is confirmed for ribbon-based devices with pristine antidots in Fig. \ref{fig_barrier_inf}, where we compare the transmission through infinite GAL ribbons (where the lead and device regions both contain identical antidot arrays) and finite GAL barriers (where a finite GAL ribbon device is connected to pristine GNR leads). 
All the device regions in this section are $W_R = 10$ unit cells wide, which corresponds to a $140$-ZGNR or a width of approximately $30 \mathrm{nm}$. 
The device length, $D_L$, is an integer number of unit cell widths and for the  $\{7,3.0\}_C$ case each of these is approximately $5 \mathrm{nm}$.

For the infinite GAL ribbon, the transmission curve (dashed, black line) clearly shows a gap corresponding to that in the DOS of the corresponding infinite GAL sheet, illustrated by the grey shaded area. 
Outside the gap, the transmission takes a series of integer values, as expected by the finite-width, periodic nature of the system. 
These integer values correspond to the number of quasi one-dimensional channels available at a given energy. 
The solid, red curve corresponds to the transmission through a system with clean graphene leads and $D_L = 4$, where we note the persistence of the transport gap observed for the infinite case. 
This behaviour is clear also in the LDOS and current maps presented in the bottom left panel for an energy in the bandgap ($E_F=0.1|t|$).
The LDOS, shown by the blue shading, quickly vanishes when we move into the antidot region and there are no channels available for transmission.
Outside the gap, we note that the transmission takes approximately half the value of the infinite case and this reduction in transmission emerges from scattering at the two interfaces between the device and clean graphene leads. 
These interfaces are not present in the infinite GAL ribbon case. 
Interference effects lead to additional oscillations for the finite width transmission and the oscillation frequency increases with $D_L$.
These are visible also in the non-uniform LDOS distribution shown for $E_{F}=0.2|t|$ in the bottom right panel.
The transmission through the barrier is evident from the bond current map shown by the orange arrows.
The inset of the top panel plots the transmission in log scale as a function of device length $D_L$ for a number of different energies. 
In each case the transmission is averaged over a narrow energy window of width $0.01|t|$ centered on the corresponding symbol shown on the energy axis of the main plot. 
We consider two energy values in first gap (black circle and magenta diamond), two values in the conducting region (red x and green triangle) and one value in the second gap (blue square). 
For each energy considered we note a quick convergence of the transmission as the device length is increased. 
However the number of unit cells required for convergence at bandgap energies increases with energy.
This suggests that higher order gaps present in GAL band structures may be more difficult to achieve in finite systems, even before consideration of the disorder effects which we shall address next.

\subsection{Antidot edge disorder}
\label{subsec_edge_roughness}

\begin{figure}[h]
\centering
\includegraphics[width =0.45\textwidth]{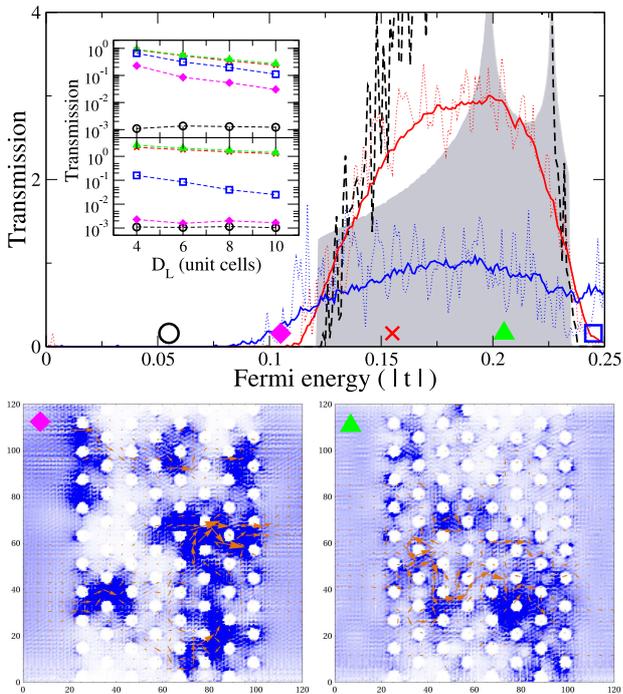}
\caption{Effect of antidot edge disorder on $D_L=4$  $\{7,3.0\}_C$ GAL barriers. The dashed black curve plots the transmission through a pristine GAL barrier, whereas the red (blue) curves represent transmission through the same system but with $\delta R_{dis} = a$ antidot edge disorder of strength $S = 0.5|t|$ ($S = |t|$) applied. The lighter, dotted curves corresponds to a single configuration whereas the heavier curves corresponds to averages over 100 such configurations. The grey shading shows the rescaled DOS of the corresponding infinite GAL sheet. The insets show the effect of altering the device length for the same energy ranges considered in Fig. \ref{fig_barrier_inf}, where now the bottom (top) panel corresponds to $S = 0.5|t|$ ($S = |t|$). The bottom panels again show LDOS and current maps at two energies as in Fig. \ref{fig_barrier_inf}, but for a specific instance of $S = |t|$ edge disorder.}
\label{fig_barrier_mild}
\end{figure}

The first disorder type we consider is the Anderson-like antidot edge roughness discussed in Sec. \ref{method_disorder}. 
The main panel in Fig. \ref{fig_barrier_mild} shows transmissions through a $D_L = 4$ GAL barrier, where the pristine  $\{7,3.0\}_C$ antidot case is shown by the dashed black curve.
The red curves correspond to transmission where an edge roughness of radius $\delta R_{dis} = a$ and strength $S = 0.5|t|$ has been applied. 
The light, dotted curve represents the transmission for a single configuration and the heavier dashed curve is the average over $100$ such configurations. 
The configurational averaging removes the configuration-dependent oscillations and reveals the overall trend for this type of disorder. 
In this case, most of the transport gap has been preserved while the transmission in the conducting region is considerably reduced compared to the pristine system. 
The gap size is reduced slightly by a broadening of the conducting region. 
The dependence of these features on the device length is demonstrated in the bottom panel of the inset, where, similar to Fig. \ref{fig_barrier_inf}, the transmissions are plotted as a function of $D_L$ for the same energy ranges. 
For the energies in the first gap (circles and diamonds) we note the robustness of the transport gap and only extremely minor fluctuations of the miniscule transmission. 
For the energies in the first conducting range, the appearance of a linear decrease in transmission in this logarithmic plot corresponds to an exponential decay with device length and hence the introduction of localisation features. 
Interestingly, this type of behaviour is also seen for the blue squares corresponding to energies in the second gap of the bulk GAL. 
This suggests that the introduction of disorder establishes a non-zero conductance in this energy range for narrow barriers, which then decays due to localisation effects as the barrier width is increased. 

The blue curves in the main panel of Fig. \ref{fig_barrier_mild} correspond to single configuration and averaged transmissions for a slightly stronger disorder ($S = 1.0|t|$).
We note that the features discussed for the $S = 0.5|t|$ case above are now even more prominent, with significant extra broadening and quenching in the conducting region. 
In fact the broadening of the conducting region has now removed the second gap completely, and gives rise to transmissions at energies far from the band edge of the pristine or milder disorder cases. 
This is confirmed by the $D_L$ dependence plotted in the upper panel of the inset.
We note that the magenta diamonds, corresponding to energies in the gap previously, are now also displaying signatures of localisation effects and significant non-zero transmission for shorter device lengths.
Examining the LDOS map for such a disorder at $E_F=0.1|t|$ (bottom left panel) shows that the density of states no longer vanishes uniformly throughout the barrier as in the pristine case.
Instead we observe that clusters of finite DOS exist within the panel region.
Breaking the perfect periodicity of the antidot lattice leads to the formation of defect states whose energies can lie in the bandgap of the pristine GAL.
The formation of such states localised at absent antidots in an otherwise pristine lattice has previously been studied\cite{Pedersen:GALscaling}. 
Similar defect states occur around each disordered antidot in the system under consideration here, and the coupling between such states gives rise to the larger clusters.
The overlap of these clusters provides percolation paths through the device region\cite{percolation_footnote}, shown by the orange arrows mapping the averaged bond currents, leading to finite transmissions. 

\begin{figure}
\centering
\includegraphics[width =0.45\textwidth]{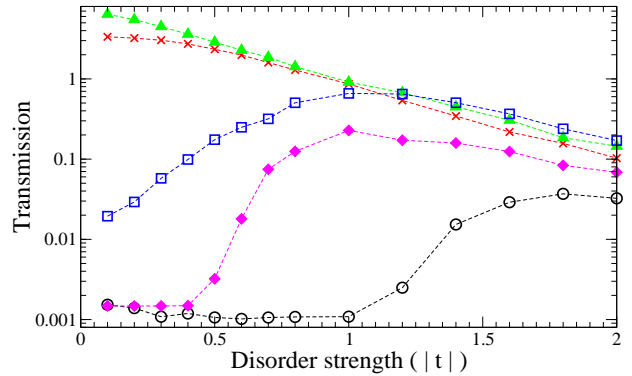}
\caption{Averaged transmissions through a $\{7,3.0\}_C$ GAL barrier as a function of antidot disorder strength $S$ for energy ranges in the first bandgap (black circles, magenta diamonds), conducting region (red Xs, green triangles) and second bandgap (blue squares) of the pristine GAL.}
\label{fig_strength}
\end{figure}

Within the conducting energy range, increasing the edge disorder strength for a fixed device length reduces the  transmission.
This is clearly seen from a comparison of the red and blue plots in the main panel of Fig. \ref{fig_barrier_mild}, and is associated with a corresponding decrease in the localization length. 
Fig. \ref{fig_strength} plots the averaged transmissions, for the same energy ranges considered in Figs \ref{fig_barrier_inf} and \ref{fig_barrier_mild}, as a function of the disorder strength $S$.
The intuitive behaviour of the conducting range energies is demonstrated by the monotonic decreases observed in the red (X) and green (triangle) curves. 
The curves corresponding to energies in the first (black circles, magenta diamonds) and second (blue squares) bandgaps of the pristine system are more complex. 
Here the system is initially insulating, but a finite transmission slowly increases with disorder before reaching a maximum at some critical value of disorder strength. 
Beyond this value, the transmission decreases in a similar manner to the conducting range energies. 
This non-monotonic behaviour represents an interplay between two effects.
Firstly, increasing the disorder strength leads to more overlap between the finite DOS clusters and a larger number of possible paths through the barrier.
However, scattering induced by the disorder also acts to reduce the transmission as the device length or scatterer strength is increased. 
At energies where the pristine system is conducting, shown in the bottom right panel for $E_F=0.2|t|$, the introduction of disorder does not introduce new paths through the device, and only acts to scatter electrons in the existing channels.

\subsection{Geometric disorders}
\label{subsec_geo_dis}

\begin{figure}
\centering
\includegraphics[width =0.48\textwidth]{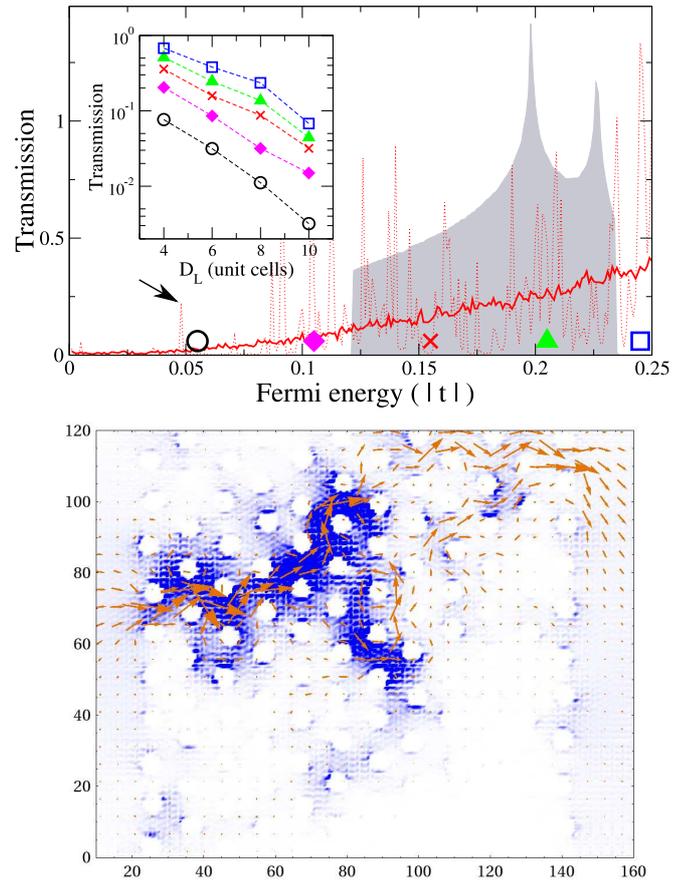}
\caption{The top panel shows single configuration (dotted) and averaged (solid) transmissions through a $D_L=6$ barrier of $\{7,3.0\}_C$ antidots with $\delta_{xy} = a$ position disorder. Grey shading shows the corresponding pristine GAL density of states. The inset shows the dependence on device length for the usual energy ranges. The bottom panel maps the LDOS (blue shading) and current distribution (orange arrows) through a disordered sample, at the energy shown by the arrow in the top panel.}
\label{fig_pos73}
\end{figure}

Until now we have not significantly altered the geometry of the antidots in the device region. 
However, even introducing disorder only at the antidot edges was capable of inducing a significant deterioration in the performance of the barrier device. 
We shall now consider the geometric disorders introduced in Section \ref{model_shapes}, where the positions and sizes of the antidots have random fluctuations. 
The red curves in the top panel of Fig. \ref{fig_pos73} show the transmission for a single instance (dotted) and a configurational average (solid) of antidot position disorder, with $\delta_{xy} = a$. 
The transmission for a single disorder instance contains many peaks, at energies both inside and outside the gap of the pristine system. 
The bottom panel maps the LDOS and current through the device region for the energy highlighted by the arrow in the top panel. 
This peak corresponds to electrons propagating through a large connected cluster of finite DOS in the device region to the opposite lead. 

The shape, size and transmission energies of these clusters are strongly dependent on the specific disorder configuration.
This leads to a smoother curve when the configurational average is taken, and this curve has no large peaks but increases reasonably steadily with Fermi energy.
The definition of a bandgap becomes difficult, as the lowest energy transmission peak occurs at different energies for different configurations. 
The inset of the top panel examines the length dependence over the usual averaged energy ranges.
We note that all energies display exponential decay behaviour, and none show the flat line several orders of magnitudes lower that was associated with bandgap energies for weak edge disorder. 
This suggests that the barrier is potentially leaky for all energies, but that reasonable on-off ratios could be established by taking advantage of the steady rise in average transmission with Fermi energy. 
This results from the greater density of single-configuration transmission peaks as the energy is increased. 
The geometry of the device, and in particular the aspect ratio, may also play a significant role. 
For a given device length $D_L$, the probability of opening a propagating channel through the device will increase with the device width $W_R$ leading to a greater number of peaks in the transmission. 

\begin{figure}
\centering
\includegraphics[width =0.48\textwidth]{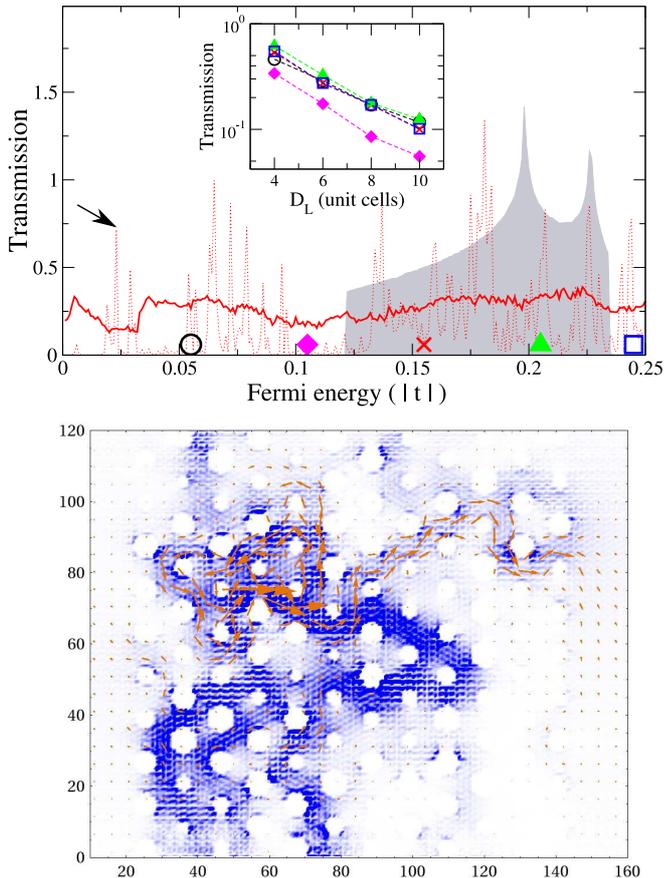}
\caption{The same barrier setup as Fig. \ref{fig_pos73}, but now with $\delta_{r} = a$ radial disorder, with the dependence on device length in the inset. Grey shading again shows the corresponding pristine GAL density of states. The bottom panel maps the LDOS (blue shading) and current distribution (orange arrows) through a disordered sample, at the energy shown by the arrow in the top panel.}
\label{fig_radial73}
\end{figure}

However, the prospects for a useable device composed of such antidots are dashed further when we examine the case of radial disorder ($\delta_{r} = a$) in Fig. \ref{fig_radial73}.
We note that the single instance and configurationally averaged transmissions display radically different behaviour from the position disorder cases. 
A higher distribution of peaks in the single configuration transmission is noted at lower energies, and there is less variation in peak heights as a function of energy. 
This leads to an average that is reasonably flat over the energy range considered here, which is reflected in the length dependence plots which mostly lie on top of each other. 
Exponential decay, signifying localisation behaviour, is again seen for all the energy ranges considered.
An increased transmission at lower energies is to be expected in a system with some distribution of smaller antidots, due to the bandgap scaling linearly with the antidot radius in pristine lattices. 
The system also contains a certain distribution of larger antidots, from which we could expect some quenching of transmissions at higher energies. 
The interplay of these two effects may partially account for the flattening of the averaged transmission as a function of energy.
The LDOS and current are mapped in the bottom panel for a low energy peak, and we note that at this energy there is a finite DOS throughout much of the device region. 

The effects of geometric disorder, and radial disorder in particular, on the bandgaps of these small antidot devices suggest that atomic levels of precision are required.
This places a huge constraint on experimental fabrication and poses a severe challenge for mass production of such devices.

\subsection{Effects of antidot size and geometry}
\label{subsec_edge_effects}

\begin{figure*}[t]
\centering
\includegraphics[width =0.99\textwidth]{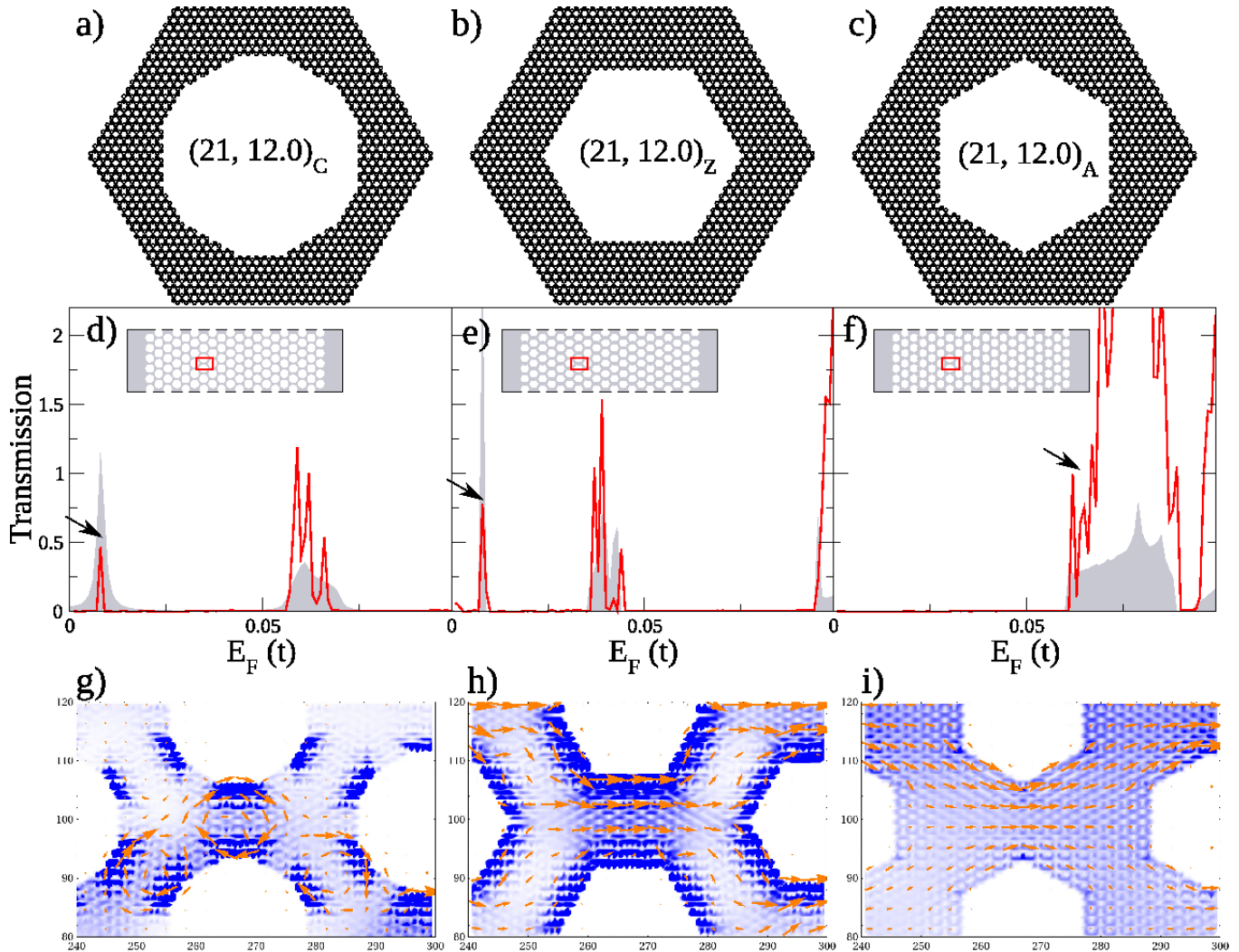}
\caption{a)-c) Unit cell geometries, d)-f) transmissions (red lines) through pristine $D_L=10,\, W_R=6$ barriers and g)-i) zoomed LDOS (blue shading) and current (orange arrows) map sections for circular, and zigzag and armchair edged hexagonal antidots. The insets in d)-f) show a schematic of the barrier devices, with the red boxes highlighting the areas shown in  g)-i). The arrows show the energies at which the maps have been calculated, and the shaded grey areas show the (rescaled) densities of states for the corresponding infinite GALs.}
\label{fig_large_geo}
\end{figure*}

In the previous section we demonstrated the drastic effect that geometric disorder has on the transport gap in a GAL barrier device. 
We focused on the $\{7,3.0\}_C$ system due to the large bandgap predicted for such a geometry. 
However, one drawback of such small, tightly-packed antidots is that even the minimum possible fluctuations in antidot radius or position constitute a significant perturbation of the system. 
At the atomic level, the position and radial fluctuations considered above give rise to situations where the total number of carbon atoms added or removed due to disorder constitute $\sim10\%$ of the total number of atoms in the pristine device region. 
Furthermore small changes in the antidot radius have a dramatic effect on the relative bandgaps of pristine GALs. 
For example, the pristine GAL composed of the largest allowed antidots in Fig \ref{fig_radial73} ($R=4a$) would have twice the bandgap expected for a GAL composed of the smallest ($R=2a$). 
We should thus expect the extreme behaviour noted for antidot lattices with such a distribution of radii.

Lattices composed of larger antidots are thus less likely to be as dramatically effected by the same absolute fluctuations in position and radius, and are also more experimentally feasible. 
In this section, we shall consider the $\{21, 12.0\}$ family of antidot lattices. 
Whereas the $R=3.0a$ antidot was too small to display any significant edge features, we see that the circular $R=12.0a$ antidot shown in Fig \ref{fig_large_geo}(a) displays an alternating sequence of zigzag and armchair edge segments. 
To explore the role of antidot geometry further, we also consider zigzag and armchair edged hexagonal antidots as shown schematically in panels b) and c) respectively. 
The `R' index in $\{L, R\}_{A/Z}$ refers to the side length for hexagonal antidots, so that zigzag and armchair edged antidots are of a similar size, but slightly smaller, than their circular counterpart with the same indices.

The middle row panels of Fig \ref{fig_large_geo} plot the conductances (solid red lines) through $D_L=10$ barriers of the three different types of $\{21, 12.0\}$ antidots, with the barrier setup shown schematically in the insets.
The densities of states of the corresponding infinite GAL systems are shown by the shaded gray areas behind the transmission curves.
We note that the simple scaling law proposed in Ref [\onlinecite{Pedersen:GALscaling}] suggests that these systems should have a bandgap of approximately $0.3 \mathrm{eV}$, so that we should expect non-zero conductance features above $\sim 0.05 |t|$. 
However, for the zigzag and circular cases there are conductance and DOS peaks at significantly smaller energies. 
Such features have been observed previously for similar systems and are associated with zigzag edge geometries\cite{Tue:finiteGAL, Wimmer2010}. 
They are related to the zero-energy DOS peaks seen in zigzag nanoribbons, but here they have their degeneracy broken and are shifted away from zero energy due to hybridizations between edges around antidots and between the edges of neighbouring antidots. 
Indeed, due to alignment between zigzag edged antidots and the hexagonal unit cell of the triangular lattice used throughout this work (and the underlying graphene lattice), we observe that pristine $\{L, R\}_Z$ systems essentially consist of a network of $2(L-R)$-ZGNRs meeting at triple junctions. 
The low energy peak features are present not only in the $\{21, 12.0\}_Z$ system, but in the $\{21, 12.0\}_C$ system also due to the alternating sequence of zigzag and armchair edges in the circular antidot. 
In contrast, the $\{21, 12.0\}_A$ system shows no low-energy peaks and by construction contains no zigzag edge segments with the possible exception of very short sections where two armchair edges meet. 
The armchair-edged antidot is also misaligned with respect to the hexagonal unit cell of the GAL, so that the lattice is not a simple network of regular width AGNRs but a more complex system which for large antidots converges to a network of connected triangular quantum dots.

The bottom panels show the density of states (blue shading) and current profile (orange arrows) for a small section (red rectangle in middle panel insets) of each device, at the energies highlighted by the arrows in the middle panels. 
The circular and zigzag antidot cases both correspond to a low-energy peak, and confirm that these low-energy transport channels are mediated by states principally localised near zigzag geometry edge segments which are shaded dark blue in the density of states plots. 
The current behaviour in these two cases is quite different however due to the formation of circulating current patterns between zigzag segments of neighbouring antidots in the $\{21, 12.0\}_C$ case. 
This is in contrast to the more uniform current behaviour noted for the zigzag-edged antidots.
The narrow band of peaks present at slightly higher energies (just above (below) $0.05 |t|$ for circular (zigzag) antidots) is also quite strongly localised near zigzag edges and gives rise, in the $\{21, 12.0\}_C$ case, to currents circulating around individual antidots. 
As in the case shown here the current loops are not entirely self contained, and electrons propagate between neighbouring loops to provide a transport channel through the barrier.
These low energy channels through antidot barriers with zigzag edge segments significantly reduce the effective transport gap of the devices. 
Furthermore, the first bandgap in these devices emerges from the splitting of the zero-energy peak associated with an infinite zigzag edge.
This is separate to the splitting of these states which may occur due to electron-electron interactions and which may lead to spin-polarised zigzag edge states\cite{Trolle-spinGAL}, similar to those predicted for ZGNRs.
A study of spin-polarised systems is beyond the scope of the current work, but it is worth noting that for antidot geometries consisting of long narrow ZGNRs (i.e. large $L$ and $R$ values) the bandgap is predicted to depend only on the ribbon width\cite{Trolle-spinGAL} and the usual bandgap scaling law\cite{Pedersen:GALscaling} no longer holds.
It is worth noting, however, that disorder is predicted to interfere significantly with the formation of spin-polarised edge states.\cite{Kunstmann2011}
This suggests that robust bandgaps as a result of spin splitting may only be a feature of pristine lattices with particular geometries.

Larger zigzag antidots, which tend to have longer zigzag segments, will have smaller bandgaps within our spin-unpolarised model due to weaker coupling with other nearby zigzag segments.
This is in direct opposition to the simple scaling behaviour of smaller antidots, and suggests that zigzag edge segments may be unsuitable for applications involving sizeable bandgaps even before we consider the introduction of disorder. 
In contrast, the $\{21, 12.0\}_A$ system has a bulk density of states and barrier transmission properties, shown in Fig \ref{fig_large_geo}f), which are more consistent with the simple scaling law. 
There are no sharp peaks in the gap region, and the LDOS map and current profiles corresponding to a bulk conducting energy, reveal a more uniform behaviour than the other antidot types and an absense of edge effects.
In the pristine case at least it appears that armchair-edged antidots have a significant advantage over zigzag or mixed edges for applications requiring a transport gap.

\begin{figure}
\centering
\includegraphics[width =0.48\textwidth]{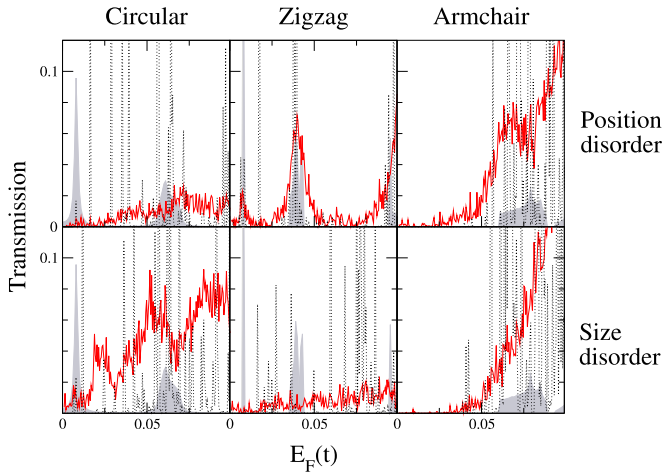}
\caption{Single instance (black, dashed lines) and configurationally averaged (red lines) transmissions through $\{21, 12.0\}$ barriers composed of circular (left), zigzag (centre) and armchair (right) edged antidots with $\delta_{xy}=a$ position disorder (top row) or  $\delta_{r}=a$ size disorder. Grey shading shows the DOS of the corresponding pristine GAL sheet.}
\label{fig_geo_dis}
\end{figure}

\begin{figure}
\centering
\includegraphics[width =0.48\textwidth]{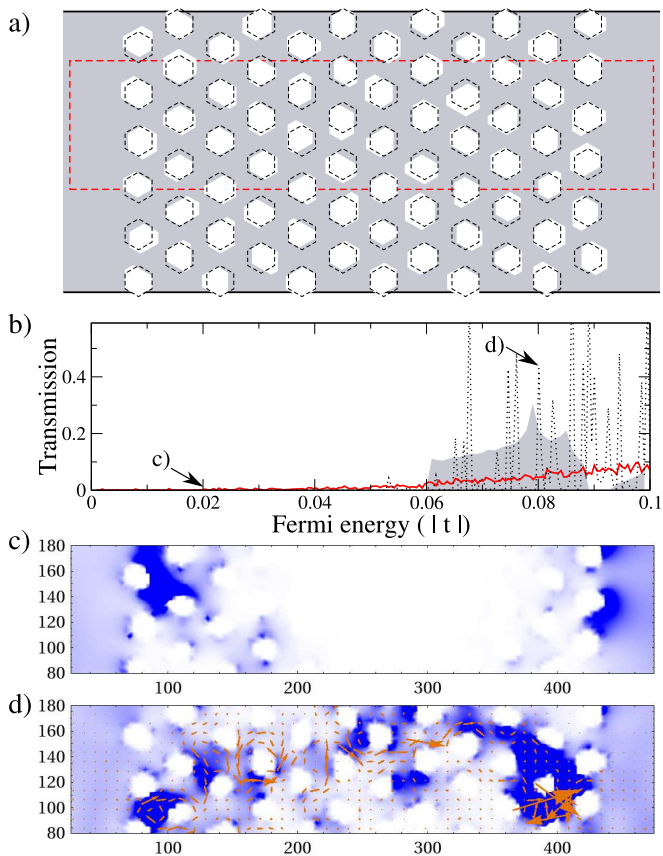}
\caption{A barrier of $\{21, 12.0\}_A$ antidots with strong position ($\delta_{xy}=3a$) and radial ($\delta_{r}=3a$) disorders applied, in addition to an edge roughness disorder of strength $S=|t|$. The pristine case is shown by dashed black lines in a). The disorder instance shown in a) has the tranmission shown by the black dotted curve in b), alongside an average (red) over 100 such configurations and the DOS of the pristine $\{21, 12.0\}_A$ sheet (grey shading). The LDOS / current maps in c) and d), shown for the red dashed region in a), show that this system is still an effective barrier at energies in the bandgap of the pristine GAL.}
\label{fig_ac_verydis}
\end{figure}

The panels in Fig. \ref{fig_geo_dis} show the effects of position and size disorder on $D_L=6$ barrier devices for each antidot geometry considered in the upper panels of Fig \ref{fig_large_geo}. 
We note that radial, or size, disorder is applied slightly differently to hexagonal antidots.
In this case, the fluctuation is applied separately to each edge so that its perpendicular distance from the antidot centre is modified by an amount in the range $[-\delta_r, \delta_r]$.
The size disorder for these antidots thus removes the regularity of the hexagons and results in zigzag/armchair segments of varying lengths within the one antidot.
This disorder thus mimics growth or treatment methods which may favour the formation of a particular edge geometry, but without necessarily resulting in perfectly regular perforations.
An example of armchair-edged GALs with such a disorder can be seen in the upper panel of Fig \ref{fig_ac_verydis}.

The red curve in each panel of Fig. \ref{fig_geo_dis} shows the transmission averaged over $200$ instances of disorder and the dotted black line shows the transmission for a single configuration. 
The gray shading once more shows the DOS for an infinite pristine GAL.
The strength of the disorder is again $\delta_{xy}=a$ for position disorder and $\delta_r=a$ for size disorder.
The behaviour in these plots can be compared to that observed for the same disorder types in smaller antidots shown in the upper of Figs. \ref{fig_pos73} and \ref{fig_radial73}. 
For all three antidot geometries the magnitude of the transmission is significantly smaller than for the pristine cases due to localisation effects. 
However, the most striking feature in these plots is once more the contrast between the armchair-edged hexagonal and the other antidot types.
The single configuration curves for circular and zigzag antidot barriers contain many peaks throughout both the bandgap and conducting energies of their pristine counterparts. 
This is because these disordered systems contain percolating transport channels between zigzag segments which occur at energies near the pristine channel energies shown in Fig \ref{fig_large_geo}.
Averaging over many configurations, we can see either broadening (as in the zigzag case with position disorder) or a complete smearing of the pristine features.
The picture is very different for the armchair case, where disorder induces merely a broadening of the pristine conducting range, so that the bandgap is reduced but not removed.
The effect seen here is similar to that of mild edge disorder in the smaller antidot systems shown in Fig. \ref{fig_barrier_mild}.
This is not surprising as the length scale of geometric disorder considered here is small compared to the antidot size and separation so that it behaves effectively as an edge perturbation.

In Fig. \ref{fig_ac_verydis} we examine the transport properties of $\{21, 12.0\}_A$ barriers with much stronger levels of disorder. 
Panel a) shows a barrier setup where both a position disorder ($\delta_{xy}=3a$) and a size disorder ($\delta_r=3a$) have been applied. 
In addition, an edge roughness disorder ($S=|t|$) has been applied to atoms within $\delta R_{dis} = a$ of the new antidot edges.
The transmission for this disorder realisation, shown in panel b) by the black dotted curve, mainly consists of a series of peaks which occur above the band edge of the pristine GAL.
The configurationally-averaged transmission (solid, red curve) suggests that the transmission remains near zero throughout much of the pristine bandgap for every configuration, before configuration-dependent peaks lead to a finite averaged transmission above the band-edge.
Unlike the mild geometric disorders considered for zigzag and circular antidots earlier, there is a clear contrast between the transmission behaviour above and below the band edge, and this is visible in both the single and averaged transmission plots.
Panels c) and d) show the LDOS and current maps for the red-dashed region highlighted in panel a), for the energies shown by arrows in panel b).
It is clear that the barrier is still effective at the lower energy, and that transmission through the disordered barrier can be switched on by raising the Fermi energy.
In fact, this highly disordered barrier proves a more robust switch than any of the circular, zigzag-edged or $\{7, 3.0\}_C$ antidot systems with much milder geometric disorder. 

These results suggest that fabrication methods which favour the formation of armchair edged perforations are key to producing antidot lattices whose electronic features are robust in the presence of mild to medium strength geometric disorder.

\section{Finite GAL waveguides}
\label{sec_WG}

\begin{figure}
\centering
\includegraphics[width =0.40\textwidth]{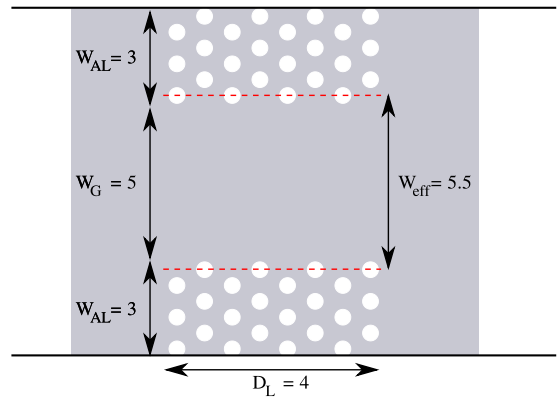}
\caption{Schematic of a finite GAL waveguide. Shown are the device length $D_L$ and the width of the antidot ($W_{AL}$) and clean graphene ($W_{G}$) strips across the ribbon. The \emph{effective width} of the waveguide $W_{eff}$, used for fitting to analytic models and in defining current confinement to the waveguide region, is also shown.} 
\label{fig_wg_schem}
\end{figure}
We now turn our attention to the waveguide geometry, where GAL regions surrounding a pristine graphene strip are employed to confine electron propagation to quasi-one dimensional channels in the strip. 
The formation of these propagating channels has been investigated previously for infinite waveguides with small, circular antidots and the band structure of such systems was calculated using both numerical and analytic, Dirac equation based approaches. \cite{GALWG}
Waveguiding in graphene has also been investigated using gates, instead of antidots, to define the waveguide edge and induce confinement.\cite{Williams2011}
Aside from immediate device application, graphene waveguides may provide a platform for exploring fundamental physical phenomena like Coulomb drag. \cite{Shylau2014}
A schematic of a finite GAL waveguide is shown in Fig. \ref{fig_wg_schem}, where the device length $D_L$ is defined as in the barrier case. 
The total ribbon width is now $$W_R = 2 W_{AL} + W_G\,,$$ where $W_{AL}$ counts the number of rectangular cells of antidots surrounding $W_G$ pristine graphene cells.
We also define the \emph{effective width} of the waveguide, from Ref. \onlinecite{GALWG}, as $$W_{\rm eff} = W_G + \tfrac{1}{2}\,.$$
This parameter takes into account that the propagation channels through an infinite pristine waveguide do not decay immediately at the edge of the rectangular unit cell, but survive a small distance into the antidot region.
It has been used to successfully describe the band structure of GAL waveguides using a gapped graphene model in Ref. \onlinecite{GALWG}.
In this work we examine how the electron transport and confinement depends on the waveguide length and on the presence of disorder in the antidot regions.
We consider first the smaller $\{7, 3.0\}_C$ antidots with and without disorder before considering the role of antidot edge geometry in larger antidot systems.

\subsection{Pristine $\{7, 3.0\}_C$ GAL waveguides}

\begin{figure}
\centering
\includegraphics[width =0.45\textwidth]{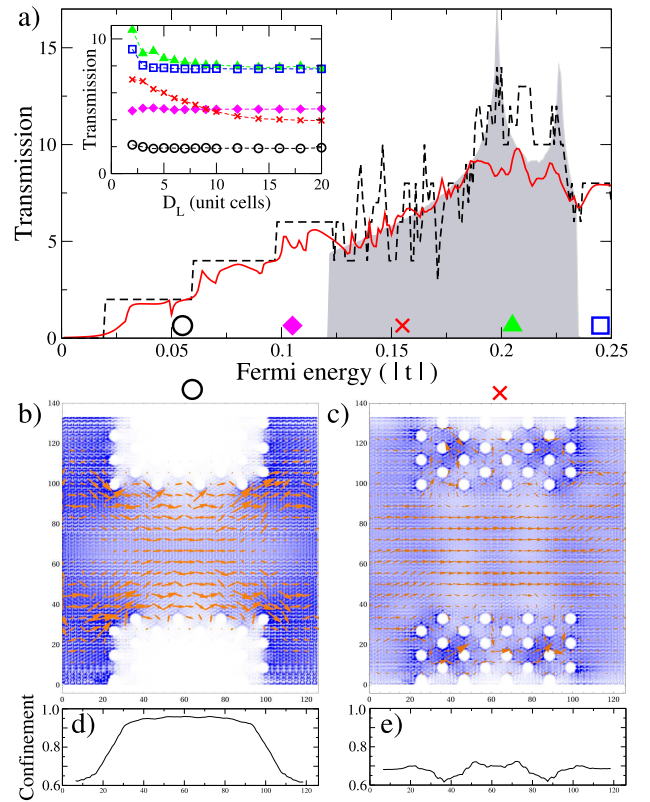}
\caption{a) Transmission through pristine $D_L=4$ (red) and infinite (black, dashed) $\{7, 3.0\}_C$ GAL waveguide systems, with the length dependence over the usual energy ranges for $\{7, 3.0\}_C$ systems shown in the inset. Grey shading shows the rescaled $\{7, 3.0\}_C$ GAL density of states. Panels b) and c) show LDOS / current maps for the $D_L=4$ waveguide at $E_F=0.05|t|$ and $E_F=0.15|t|$ respectively. Panels d) and e) show the confinement of current to the waveguide region along the length of the waveguide for the same energies.}
\label{fig_wg_length}
\end{figure}

The transmission through an infinite pristine waveguide, composed of $\{7, 3.0\}_C$ antidots with $W_{AL}=3$ and $W_G=5$, is shown by the dashed black plot in Fig. \ref{fig_wg_length} a).
We observe the onset of a series of sharp plateaux as the Fermi energy is increased within the bandgap of pristine GAL.
These correspond to transport through one-dimensional channels in the pristine graphene region.
The energy dependence and bandstructure of these channels, calculated previously,\cite{GALWG} agree with the finite-width transport calculation here.
More complex behaviour is seen above the bulk GAL band edge due to the presence of both waveguide and bulk-like states.
The red curve shows the transmission through the finite length waveguide system shown schematically in Fig. \ref{fig_wg_schem}.
The general trend of the plateaux is still clearly discernable, but the exact integer values of transmission are no longer achieved due to scattering at the interfaces between the clean ribbon leads and the waveguide device region. 
From the inset it is clear that the transmission values converge quite quickly as the device length is increased, particularly for energies in the gap (waveguiding) ranges.
As in the barrier case, we have taken averages over a small energy range in each case to smooth out device length dependent oscillations.
The LDOS / current maps in panels b) and c) of Fig. \ref{fig_wg_length}, taken at $E_F=0.05|t|$ and $E_F=0.15|t|$ respectively, illustrate clearly the efficiency of the waveguide at energies in the bandgap of the pristine GAL, and its lack thereof at energies in which the pristine GAL is conducting.
Each GAL section acts as a barrier in panel b), effectively `funneling' electrons into the waveguide region. 
This role is no longer performed as the Fermi energy is increased and electrons pass freely into the GAL sections, as shown in panel c).

To examine this feature further, we define the \emph{confinement} of current to the waveguide region. 
The magnitudes of the bond currents are summed across the width of the device for a fixed distance along its length.
The confinement is then defined as the fraction of this quantity that lies within the waveguide's effective width $W_{\rm eff}$, as illustrated in Fig. \ref{fig_wg_schem}.
The confinement, as a function of position along the device length, at $E_F=0.05|t|$ and $E_F=0.15|t|$ is plotted in Fig. \ref{fig_wg_length} d) and e), where a small spatial averaging has been applied to remove oscillations on length scales less than the width of an antidot.
For the lower energy the confinement increases sharply at the edge of the waveguiding region, so that within one unit cell or so approximately 95\% of the current is passing through the waveguide region.
The opposite effect is seen at the far end of the device as the current diffuses once more across the width of the pristine graphene lead.
This matches the behaviour seen in the corresponding LDOS / current map.
For the higher energy, in the conducting energy range of the pristine GAL, only minor fluctuations in the confinement are seen across the length of the device.
This again corresponds closely to the behaviour observed in the associated map where current runs in both the antidot and graphene regions.

These results confirm that, similar to the barrier case, only a small number of pristine antidot rows are required for waveguiding systems to achieve similar results to those predicted for infinite, periodic systems.

\subsection{Disorder and geometry effects}

\begin{figure}
\centering
\includegraphics[width =0.45\textwidth]{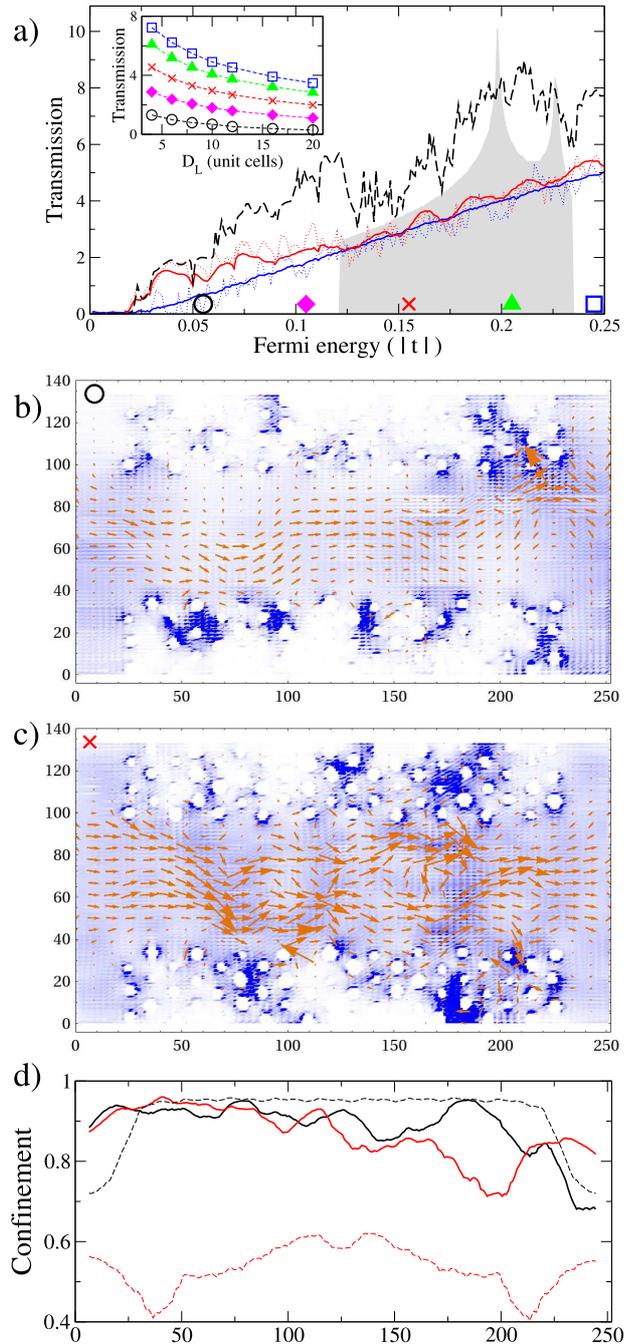}
\caption{a) Transmission through a $D_L=10$ waveguide composed of pristine $\{7, 3.0\}_C$ antidots (black, dashed) or with $S=|t|$ edge roughness (red) or  $\delta_{xy}=a$, $\delta_{r}=a$ geometric disorder (blue) applied. The dotted curves represent single configurations and the solid curves configurational averages. Grey shading shows the bulk GAL density of states. The inset shows the dependence on device length for different energies for the geometric disorder case. Panels b) and c) show LDOS and current maps though a geometrically disordered sample at $E_F=0.05|t|$ and $E_F=0.15|t|$ respectively.
d) Solid lines show the confinement along the length of the device for the energies in b) (black) and c) (red). The dashed lines show the confinement for the corresponding pristine systems.}
\label{fig_wg_dis}
\end{figure}

Since the waveguide system is essentially composed of two antidot barrier regions we might expect to see similar behaviour in disordered waveguide devices to that noted earlier for disordered barriers.
An important distinction, however, is that we are now primarily interested in transmission through \emph{pristine} graphene regions bordered by GAL regions, and not necessarily through the GAL regions themselves.
In the main panel of Fig. \ref{fig_wg_dis}a) we plot the transmission through a $D_L=10$ waveguide composed of $\{7, 3.0\}_C$ antidots with either an $S=|t|$ edge disorder (red curves) or a combination of $\delta_{xy}=a$ position disorder and $\delta_{r}=a$ size disorder (blue curves).
As usual dotted curves represent single configurations and the solid lines configurational averages.
The geometric disorders applied individually (not shown) display qualitatively similar behaviour throughout the waveguiding and GAL conducting energy ranges, with size disorder only resulting in larger conductances at very low energies before the onset of the first waveguiding plateau.
The strong antidot edge roughness (red) leads to some suppression of the transmission, but the onset of many individual waveguide plateaux is still visible although their features have been smoothened.
In constrast, the averaged transmission for geometric disorder (blue) increases uniformly as the Fermi energy is increased and the plateaux features have been completely removed.
Interestingly we note that, unlike the pristine case, the behaviour for both disorder types is not particularly dependent on whether we are in the waveguiding or GAL conducting energy regimes.
This is also clear in the device length dependence shown in the inset, where we note similar behaviour for all the energies considered. 
The transmission in each case decays far slower than for the barriers examined in Figs. \ref{fig_pos73} and \ref{fig_radial73}.

This uniform behaviour across different energies arises because the type of disorder we are applying strongly suppresses transmission through the GAL regions, as we saw for the barrier devices discussed previously.
However, it acts more like an edge disorder to channels propagating through the waveguide region.
Finite LDOS clusters in the GAL regions act as resonant scatterers to electrons in the low energy waveguide channels, leading to dips in the transmission for specific energies and disorder realisations.
These dip features tend to be averaged out over many disorder realisations.
Extended clusters can lead to transport channels in the GAL regions which may either rejoin the waveguide or cause leakage to external regions.
The magnitude of the leakage transmission will decay exponentially with $W_{AL}$, the width of the antidot region, in a manner similar to the transmission through geometrically disordered barriers.
However, the majority of the transmission is carried within the waveguide region, as shown in the LDOS / current map of Fig. \ref{fig_wg_dis}b) for a low energy ($E_F=0.05|t|$).
Transmission in pristine infinite waveguides at energies above the GAL band edge displays a complex non-monotonic behaviour due to the interplay between waveguide channels and bulk GAL states, as seen in the dashed plot in Fig. \ref{fig_wg_length}.
Geometric disorder strongly suppresses the contribution of bulk GAL states, whilst continuing to scatter waveguide states in a similar manner to those at lower energies.
Thus an interesting consequence of introducing disorder is that we increase the energy range over which transmission through the device is predominantly mediated by waveguide channels.
This is evident in the map of Fig. \ref{fig_wg_dis}c), where we plot the LDOS / current map for a higher energy ($E_F=0.15|t|$).
For both energies we note that the current is mainly contained in the waveguide region, but that the current pattern is far more non-uniform than for the waveguiding channels in pristine devices. 
The confinement of electrons through the disordered device at both energies is plotted in  Fig. \ref{fig_wg_dis}d), alongside their pristine counterparts (dashed lines). 
The disordered device still displays confinement comparable to the pristine system at low energies (black lines), whereas in the high energy case (red lines) the confinement is considerably improved by the introduction of disorder, although we do note significant reductions at points associated with leakage paths visible in Fig. \ref{fig_wg_dis}c).

\begin{figure}
\centering
\includegraphics[width =0.45\textwidth]{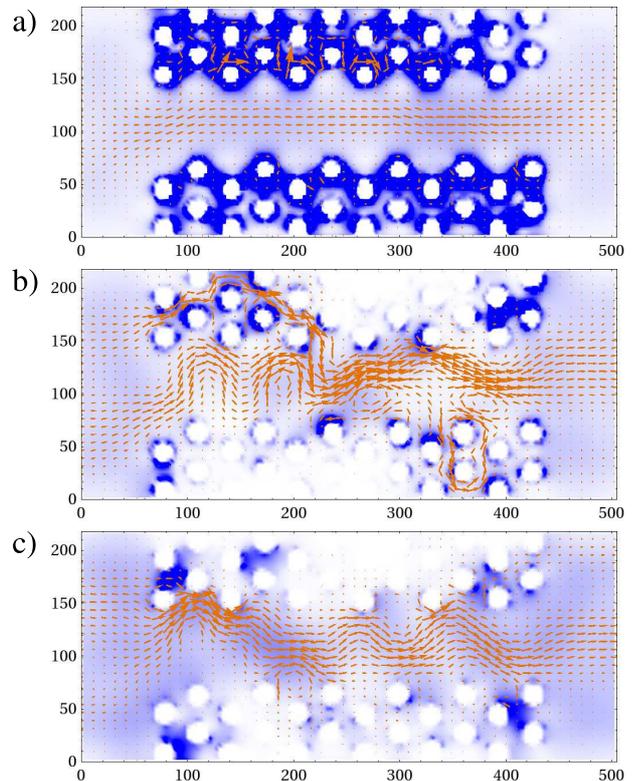}
\caption{LDOS and current maps for: a) A pristine  $\{21,12.0\}_Z$ waveguide at $E_F=0.041|t|$. At this energy poor confinement is seen due to the presence of antidot edge states.
b) A $\{21,12.0\}_Z$ waveguide with geometric disorder at $E_F=0.050|t|$ where very poor confinement is seen and leakage channels mediated by edge states are clearly visible. The pristine waveguide at this energy shows excellent confinement.
c) A $\{21,12.0\}_A$ waveguide with geometric disorder at $E_F=0.052|t|$, where leakage channels reduce confinement but are not mediated by edge states.}
\label{fig_wgbig_dis}
\end{figure}

The performance of waveguide devices appears to be less dependent on atomic level precision than that of barrier devices.
This is because although disorder can introduce some degree of leakage, it in general inhibits the flow of electrons in the GAL regions bordering the waveguide.
A bandgap in these regions, emerging from either periodicity-induced confinement or disorder-induced localisation effects, acts as a funnel for electrons into the waveguide region.
It is now worth determining what role the additional zigzag edge channels that appear for larger antidots, and which hindered the performance of both pristine and disordered barriers, play when incorporated into waveguide device geometries.
Due to the large system sizes involved, we restricted our calculations to waveguide devices with $W_{AL}=W_G=2$ for the $\{21,12.0\}_{C/Z/A}$ antidots considered.
The reduced number of antidot rows defining the waveguide meant that leakage channels opened very quickly for all antidot types once geometric disorders were introduced.
Nonetheless, the low energy states associated with zigzag edges were seen to introduce additional features for waveguides with circular or zigzag hexagonal edge geometries.
Some example LDOS / current maps for zigzag and armchair antidot waveguides are shown in Fig \ref{fig_wgbig_dis}.
In pristine zigzag antidot devices, reduced confinement was observed at the energies associated with the edge states (panel a).
At other energies in the waveguiding range a very efficient confinement of electrons to the waveguide was seen.
Similar features were noted for large circular antidots containing zigzag edge segments.
However, as in the barrier case, the introduction of geometric disorder significantly increases the energy range over which zigzag edge states mediate leakage paths in the antidot regions.
This is seen clearly in the map in panel b), which is taken at an energy for which the pristine $\{21,12.0\}_Z$ waveguide provides excellent confinement.
Waveguides composed from pristine $\{21,12.0\}_A$ antidots are very efficient until the band edge of the associated bulk GAL.
The disordered armchair cases do contain leakage paths, but these are not mediated by antidot edge states and reasonable confinement is seen for most energies, as shown in the example in panel c).
The behaviour of armchair barrier devices suggests that these disordered waveguides will perform better for larger values of $W_{AL}$, which are beyond the reach of the present simulations, but which should quench leakage currents more thoroughly.
We should thus expect larger scale armchair-edged devices to have a similar performance to the $\{7,3.0\}_C$ waveguides discussed earlier, whereas antidots with zigzag edge segments will be less efficient at confining electrons to the waveguide channels once disorder is included.

\section{Conclusions and discussion}
\label{sec_conclude}
In this work we have examined a wide range of devices constructed from graphene antidot lattices with different geometries and with different types of disorder applied.
The main device type we considered consisted of a graphene nanoribbon with a finite-length GAL barrier across the width of the ribbon.
For small, atomically pristine antidots we confirmed that such a configuration acted as an effective switch which blocked the flow of electrons with energies in the bandgap of the associated antidot lattice.
However these small, tightly packed antidots, which give the largest bandgaps in the pristine case are also those most strongly affected by small amounts of disorder.
While the switch retained a large degree of efficacy for mild antidot edge roughness, the inclusion of minimal amounts of geometric disorder completely removed the required behaviour as the barriers leaked heavily at low energies and higher energy transmissions were suppressed by scattering.

Larger antidots, although expected to have smaller bandgaps, should be more robust against the same absolute levels of geometric disorder.
However, for these systems we discovered that the edge geometry of the individual antidots played an important role.
Large circular antidots have significant zigzag geometry segments along their edges, and these were seen to contribute low energy leakage channels through barrier devices, even for pristine systems.
The same effect was noted for antidots consisting entirely of zigzag edges.
When disorder was introduced, the energy range over which zigzag-edge mediated leakage channels appeared was considerably broadened, leading to particularly ineffective barrier devices.
On the other hand, antidots containing exclusively armchair edge segments were found to display excellent barrier characteristics in the pristine case, and also to be far more robust in the presence of quite strong geometric disorder.
These results suggest that fabrication or treatment techniques which favour armchair edge geometries will lead to significantly better performance in barrier devices.
On the other hand, zigzag edge antidots may be useful for applications not requiring a gap, but which exploit the magnetic properties predicted by other works for such systems.
In this light, recent experimental progress in selectively etching particular edge geometries\cite{Oberhuber2013, Pizzocchero2014}, and heating-induced reconstruction of edges\cite{Jia2009, Xu2013} to form these geometries, is particularly promising.

We have also considered waveguide geometry devices, where rows of antidots were used to define strips of pristine graphene to act as electronic waveguides. 
Finite-length pristine devices were found to display similar transport characteristics to their previously investigated infinite-length counterparts.\cite{GALWG}
Disorder introduced some scattering of electrons in the waveguides, and opened up some leakage paths, but was not found to be as detrimental to device performance as in the barrier setup.
Armchair-edged antidots were found to perform better than those with zigzag or circular edges, due to similar edge-mediated leakage effects to those in barrier devices.
Reasonable performance in waveguide devices can be expected once transmission is sufficiently suppressed in the `gapped' regions defining the waveguide.
Once disorder is restricted to this region, and scattering of channels in the waveguide region itself is kept to a minimum, disorder can enhance the guiding effect by extending it to energy ranges beyond those dictated by the underlying bandgap of the associated antidot lattice. 

\begin{center}
\begin{table}
 \begin{tabular}{| c | c | c | c | c | c |}
\hline
Set & $\;\{\,L, R\,\}$\; &  $\sigma_w$ (nm) &\; $\delta_{xy/r}$ \;& $\delta_{xy} = \delta_{r}$ & \; Reference \; \\
\hline
\hline
a & $\{90, 50\}$ & $2.0$ & $10$ & $7$ & [\onlinecite{Bai2010}] Fig. 3 a/e/i\\ \hline
b & $\{90, 56\}$ & $1.9$ & $10$ & $7$ & [\onlinecite{Bai2010}] Fig. 3 b/f/j\\ \hline
c & $\{90, 65\}$ & $1.5$ & $8$ & $5$ & [\onlinecite{Bai2010}] Fig. 3 c/g/k\\ \hline
d & $\{63, 36\}$ & $1.3$ & $6$ & $4$ & [\onlinecite{Bai2010}] Fig. 3 d/h/l\\ \hline
e & $\{85, 45\}$ & \multirow{4}{*}{$< 2.0$} & \multirow{4}{*}{$< 10$} &  \multirow{4}{*}{$< 7$} & [\onlinecite{Kim2012}] Fig. 2 a\\ \cline{1-2}\cline{6-6}
f & $\{85, 55\}$ & &  &  & [\onlinecite{Kim2012}] Fig. 2 b\\ \cline{1-2}\cline{6-6}
g & $\{85, 56\}$ & &  &  & [\onlinecite{Kim2012}] Fig. 2 c\\ \cline{1-2}\cline{6-6}
h & $\{85, 59\}$ & &  &  & [\onlinecite{Kim2012}] Fig. 2 d\\ \hline
 \end{tabular}
 \caption{Details of experimental antidot geometries and levels of disorder given by the associated references. Shown are the approximate geometry in our $\{L, R\}$ notation, the experimentally reported standard deviation in the neck-width distribution $\sigma_w$, the strength of position disorder or radial disorder only ($\delta_{xy/r}$) corresponding to this in our model, and also the required strength of a combined disorder with $\delta_{xy} = \delta_{r}$. $\delta_{xy}$ and $\delta_{r}$ are in units of $a$.}
 \label{table_experiment}
 \end{table}
\end{center}

It is worth comparing the antidot geometry and disorder levels considered in this work to those reported by experiment.
Lattices of circular antidots with feature sizes on the order of $10\mathrm{nm}$ have been reported and the level of geometric disorder in these cases is given by the standard deviation ($\sigma_w$) in the distribution of neck widths between neighbouring antidots\cite{Bai2010, Kim2012}.
This quantity is easily extracted from the geometries generated in this work, and we can associate each $\sigma_w$ with a particular strength of position ($\delta_{xy}$) or radial ($\delta_r$) disorder applied either individually or in combination.
Table \ref{table_experiment} shows the approximate antidot geometry parameters and disorder strengths corresponding to the experimental data in References [\onlinecite{Bai2010}] and [\onlinecite{Kim2012}].
We note that systems tabulated here are generally much larger than those simulated in this work and have greater levels of disorder.
However, set d has feature sizes only three times larger than the $\{21, 12.0\}_C$ antidots examined in Section \ref{subsec_edge_effects}.
The relative level of disorder in this system is also similar to that studied in this work, so that our large antidot system results should hold for experimentally realisable structures.

It is also interesting to note that there is some experimental evidence of transport gaps in GAL systems\cite{Eroms2009, Kim2010, Bai2010, Giesbers2012}.
The systems investigated in these works are composed of antidot lattices which, in comparison to the systems studied here, are strongly disordered.
The effective bandgap in these systems is generally extracted from the temperature dependence of the conductance.
It should be noted that it is difficult to differentiate transport gaps emerging from periodicity-induced confinement, or from purely disorder-induced localisation effects.
Highly disordered antidot lattices essentially consist of a random network of disordered nanoribbons, and these have been shown to be particularly susceptible individually to disorder-induced transport gaps\cite{mucciolo:graphenetransportgaps, evaldsson:ribbonedgeanderson}. 
The bandgap sizes extracted from these experiments also fit reasonably well with nanoribbon bandgap scaling laws.
Furthermore, we note that only triangular lattices aligned correctly with the underlying graphene lattice are predicted to induce a bandgap for all lattice spacings.
Other lattice geometries may be metallic or semiconducting depending on the lattice spacing, and antidots with zigzag edges follow different bandgap scaling laws. 
The experimental results were based on square\cite{Eroms2009, Giesbers2012} and triangular\cite{Kim2010, Bai2010} lattices, but in the latter case it is unclear how the antidot lattice aligns with the underlying graphene lattice.
Thus it is not even clear whether the associated pristine lattices necessarily would produce a bandgap, and it is thus very difficult to conclude that the experimental bandgaps reported to date are due to the periodic modulation of graphene outlined in Ref. \onlinecite{Pedersen:GALscaling}.

Even in atomically precise GAL systems, long-range potential disorder arising from, for example, substrate impurities may play an important role in determining transport properties.
Studies of such disorder in graphene nanoribbons\cite{Wurm2012} and quantum dots\cite{Rycerz2012} have shown that the strength of intervalley scattering depends strongly on the edge geometries and underlying symmetry classes of these structures.
We expect similar effects to arise in antidot systems, and that the magnetoconductance and universal conductance fluctuations will display a strong dependence on the antidot size, shape and edge geometry.

This work has demonstrated that the electronic structure features predicted for periodically modulated graphene systems can be extremely sensitive to atomic scale imperfections and defects.
Furthermore, we have outlined a strategy to optimise the performance of antidot based barrier devices based on these features.
However, there are other features of nanostructured systems whose interplay with disorder may prove very interesting.
The magnetic properties of zigzag-edged antidots have been investigated for the pristine case\cite{Furst2009PRB, Trolle-spinGAL}, but the effect of randomised edge lengths may introduce new features by breaking the equivalency of neighbouring edges and thus the overall spin degeneracy of hexagonal antidot systems.
The effect of chemical functionalisation\cite{Ouyang2010} on graphene antidot lattices is also of keen interest, particularly in the light of recent results suggesting that the conductance variation of highly disordered perforated graphene sheets allows for gas sensing at extremely low concentrations\cite{Cagliani2014}.
Future investigation of nanostructured graphene along these lines is bound to yield many intriguing results and extend the applicability of graphene-based materials.

\hspace{1cm}

\textbf{Acknowledgements}
The Center for Nanostructured Graphene (CNG) is sponsored by the Danish Research Foundation, Project DNRF58.
We thank T. G. Pedersen, M. Settnes and P. D. Gorman for helpful comments on the manuscript.


\begin{thebibliography}{60}%
\makeatletter
\providecommand \@ifxundefined [1]{%
 \@ifx{#1\undefined}
}%
\providecommand \@ifnum [1]{%
 \ifnum #1\expandafter \@firstoftwo
 \else \expandafter \@secondoftwo
 \fi
}%
\providecommand \@ifx [1]{%
 \ifx #1\expandafter \@firstoftwo
 \else \expandafter \@secondoftwo
 \fi
}%
\providecommand \natexlab [1]{#1}%
\providecommand \enquote  [1]{``#1''}%
\providecommand \bibnamefont  [1]{#1}%
\providecommand \bibfnamefont [1]{#1}%
\providecommand \citenamefont [1]{#1}%
\providecommand \href@noop [0]{\@secondoftwo}%
\providecommand \href [0]{\begingroup \@sanitize@url \@href}%
\providecommand \@href[1]{\@@startlink{#1}\@@href}%
\providecommand \@@href[1]{\endgroup#1\@@endlink}%
\providecommand \@sanitize@url [0]{\catcode `\\12\catcode `\$12\catcode
  `\&12\catcode `\#12\catcode `\^12\catcode `\_12\catcode `\%12\relax}%
\providecommand \@@startlink[1]{}%
\providecommand \@@endlink[0]{}%
\providecommand \url  [0]{\begingroup\@sanitize@url \@url }%
\providecommand \@url [1]{\endgroup\@href {#1}{\urlprefix }}%
\providecommand \urlprefix  [0]{URL }%
\providecommand \Eprint [0]{\href }%
\providecommand \doibase [0]{http://dx.doi.org/}%
\providecommand \selectlanguage [0]{\@gobble}%
\providecommand \bibinfo  [0]{\@secondoftwo}%
\providecommand \bibfield  [0]{\@secondoftwo}%
\providecommand \translation [1]{[#1]}%
\providecommand \BibitemOpen [0]{}%
\providecommand \bibitemStop [0]{}%
\providecommand \bibitemNoStop [0]{.\EOS\space}%
\providecommand \EOS [0]{\spacefactor3000\relax}%
\providecommand \BibitemShut  [1]{\csname bibitem#1\endcsname}%
\let\auto@bib@innerbib\@empty
\bibitem [{\citenamefont {Geim}\ and\ \citenamefont
  {Novoselov}(2007)}]{riseofgraphene}%
  \BibitemOpen
  \bibfield  {author} {\bibinfo {author} {\bibfnamefont {A.~K.}\ \bibnamefont
  {Geim}}\ and\ \bibinfo {author} {\bibfnamefont {K.~S.}\ \bibnamefont
  {Novoselov}},\ }\href@noop {} {\bibfield  {journal} {\bibinfo  {journal}
  {Nature Materials}\ }\textbf {\bibinfo {volume} {6}},\ \bibinfo {pages} {183}
  (\bibinfo {year} {2007})}\BibitemShut {NoStop}%
\bibitem [{\citenamefont {Castro~Neto}\ \emph {et~al.}(2009)\citenamefont
  {Castro~Neto}, \citenamefont {Guinea}, \citenamefont {Peres}, \citenamefont
  {Novoselov},\ and\ \citenamefont {Geim}}]{neto:graphrmp}%
  \BibitemOpen
  \bibfield  {author} {\bibinfo {author} {\bibfnamefont {A.~H.}\ \bibnamefont
  {Castro~Neto}}, \bibinfo {author} {\bibfnamefont {F.}~\bibnamefont {Guinea}},
  \bibinfo {author} {\bibfnamefont {N.~M.~R.}\ \bibnamefont {Peres}}, \bibinfo
  {author} {\bibfnamefont {K.~S.}\ \bibnamefont {Novoselov}}, \ and\ \bibinfo
  {author} {\bibfnamefont {A.~K.}\ \bibnamefont {Geim}},\ }\href {\doibase
  10.1103/RevModPhys.81.109} {\bibfield  {journal} {\bibinfo  {journal}
  {Reviews of Modern Physics}\ }\textbf {\bibinfo {volume} {81}},\ \bibinfo
  {eid} {109} (\bibinfo {year} {2009})}\BibitemShut {NoStop}%
\bibitem [{\citenamefont {Nakada}\ \emph {et~al.}(1996)\citenamefont {Nakada},
  \citenamefont {Fujita}, \citenamefont {Dresselhaus},\ and\ \citenamefont
  {Dresselhaus}}]{Nakada:1996ribbons}%
  \BibitemOpen
  \bibfield  {author} {\bibinfo {author} {\bibfnamefont {K.}~\bibnamefont
  {Nakada}}, \bibinfo {author} {\bibfnamefont {M.}~\bibnamefont {Fujita}},
  \bibinfo {author} {\bibfnamefont {G.}~\bibnamefont {Dresselhaus}}, \ and\
  \bibinfo {author} {\bibfnamefont {M.~S.}\ \bibnamefont {Dresselhaus}},\
  }\href {\doibase 10.1103/PhysRevB.54.17954} {\bibfield  {journal} {\bibinfo
  {journal} {Phys. Rev. B}\ }\textbf {\bibinfo {volume} {54}},\ \bibinfo
  {pages} {17954} (\bibinfo {year} {1996})}\BibitemShut {NoStop}%
\bibitem [{\citenamefont {Ezawa}(2006)}]{ezawa:ribbonwidth}%
  \BibitemOpen
  \bibfield  {author} {\bibinfo {author} {\bibfnamefont {M.}~\bibnamefont
  {Ezawa}},\ }\href {\doibase 10.1103/PhysRevB.73.045432} {\bibfield  {journal}
  {\bibinfo  {journal} {Phys. Rev. B}\ }\textbf {\bibinfo {volume} {73}},\
  \bibinfo {eid} {045432} (\bibinfo {year} {2006})}\BibitemShut {NoStop}%
\bibitem [{\citenamefont {Brey}\ and\ \citenamefont
  {Fertig}(2006)}]{brey_electronic_2006}%
  \BibitemOpen
  \bibfield  {author} {\bibinfo {author} {\bibfnamefont {L.}~\bibnamefont
  {Brey}}\ and\ \bibinfo {author} {\bibfnamefont {H.~A.}\ \bibnamefont
  {Fertig}},\ }\href {\doibase 10.1103/PhysRevB.73.235411} {\bibfield
  {journal} {\bibinfo  {journal} {Phys. Rev. B}\ }\textbf {\bibinfo {volume}
  {73}},\ \bibinfo {pages} {235411} (\bibinfo {year} {2006})}\BibitemShut
  {NoStop}%
\bibitem [{\citenamefont {Son}\ \emph {et~al.}(2006)\citenamefont {Son},
  \citenamefont {Cohen},\ and\ \citenamefont {Louie}}]{Son:ribbonenergygaps}%
  \BibitemOpen
  \bibfield  {author} {\bibinfo {author} {\bibfnamefont {Y.-W.}\ \bibnamefont
  {Son}}, \bibinfo {author} {\bibfnamefont {M.~L.}\ \bibnamefont {Cohen}}, \
  and\ \bibinfo {author} {\bibfnamefont {S.~G.}\ \bibnamefont {Louie}},\ }\href
  {\doibase 10.1103/PhysRevLett.97.216803} {\bibfield  {journal} {\bibinfo
  {journal} {Phys. Rev. Lett.}\ }\textbf {\bibinfo {volume} {97}},\ \bibinfo
  {eid} {216803} (\bibinfo {year} {2006})}\BibitemShut {NoStop}%
\bibitem [{\citenamefont {Pedersen}\ \emph
  {et~al.}(2008{\natexlab{a}})\citenamefont {Pedersen}, \citenamefont {Flindt},
  \citenamefont {Pedersen}, \citenamefont {Mortensen}, \citenamefont {Jauho},\
  and\ \citenamefont {Pedersen}}]{Pedersen:GALscaling}%
  \BibitemOpen
  \bibfield  {author} {\bibinfo {author} {\bibfnamefont {T.~G.}\ \bibnamefont
  {Pedersen}}, \bibinfo {author} {\bibfnamefont {C.}~\bibnamefont {Flindt}},
  \bibinfo {author} {\bibfnamefont {J.}~\bibnamefont {Pedersen}}, \bibinfo
  {author} {\bibfnamefont {N.~A.}\ \bibnamefont {Mortensen}}, \bibinfo {author}
  {\bibfnamefont {A.-P.}\ \bibnamefont {Jauho}}, \ and\ \bibinfo {author}
  {\bibfnamefont {K.}~\bibnamefont {Pedersen}},\ }\href {\doibase
  10.1103/PhysRevLett.100.136804} {\bibfield  {journal} {\bibinfo  {journal}
  {Phys. Rev. Lett.}\ }\textbf {\bibinfo {volume} {100}},\ \bibinfo {pages}
  {136804} (\bibinfo {year} {2008}{\natexlab{a}})}\BibitemShut {NoStop}%
\bibitem [{\citenamefont {Pedersen}\ \emph
  {et~al.}(2008{\natexlab{b}})\citenamefont {Pedersen}, \citenamefont {Flindt},
  \citenamefont {Pedersen}, \citenamefont {Jauho}, \citenamefont {Mortensen},\
  and\ \citenamefont {Pedersen}}]{Pedersen2008}%
  \BibitemOpen
  \bibfield  {author} {\bibinfo {author} {\bibfnamefont {T.}~\bibnamefont
  {Pedersen}}, \bibinfo {author} {\bibfnamefont {C.}~\bibnamefont {Flindt}},
  \bibinfo {author} {\bibfnamefont {J.}~\bibnamefont {Pedersen}}, \bibinfo
  {author} {\bibfnamefont {A.-P.}\ \bibnamefont {Jauho}}, \bibinfo {author}
  {\bibfnamefont {N.}~\bibnamefont {Mortensen}}, \ and\ \bibinfo {author}
  {\bibfnamefont {K.}~\bibnamefont {Pedersen}},\ }\href {\doibase
  10.1103/PhysRevB.77.245431} {\bibfield  {journal} {\bibinfo  {journal} {Phys.
  Rev. B}\ }\textbf {\bibinfo {volume} {77}},\ \bibinfo {pages} {245431}
  (\bibinfo {year} {2008}{\natexlab{b}})}\BibitemShut {NoStop}%
\bibitem [{\citenamefont {F\"{u}rst}\ \emph
  {et~al.}(2009{\natexlab{a}})\citenamefont {F\"{u}rst}, \citenamefont
  {Pedersen}, \citenamefont {Flindt}, \citenamefont {Mortensen}, \citenamefont
  {Brandbyge}, \citenamefont {Pedersen},\ and\ \citenamefont
  {Jauho}}]{Furst2009}%
  \BibitemOpen
  \bibfield  {author} {\bibinfo {author} {\bibfnamefont {J.~A.}\ \bibnamefont
  {F\"{u}rst}}, \bibinfo {author} {\bibfnamefont {J.~G.}\ \bibnamefont
  {Pedersen}}, \bibinfo {author} {\bibfnamefont {C.}~\bibnamefont {Flindt}},
  \bibinfo {author} {\bibfnamefont {N.~A.}\ \bibnamefont {Mortensen}}, \bibinfo
  {author} {\bibfnamefont {M.}~\bibnamefont {Brandbyge}}, \bibinfo {author}
  {\bibfnamefont {T.~G.}\ \bibnamefont {Pedersen}}, \ and\ \bibinfo {author}
  {\bibfnamefont {A.-P.}\ \bibnamefont {Jauho}},\ }\href {\doibase
  10.1088/1367-2630/11/9/095020} {\bibfield  {journal} {\bibinfo  {journal}
  {New Journal of Physics}\ }\textbf {\bibinfo {volume} {11}},\ \bibinfo
  {pages} {095020} (\bibinfo {year} {2009}{\natexlab{a}})}\BibitemShut
  {NoStop}%
\bibitem [{\citenamefont {F\"{u}rst}\ \emph
  {et~al.}(2009{\natexlab{b}})\citenamefont {F\"{u}rst}, \citenamefont
  {Pedersen}, \citenamefont {Brandbyge},\ and\ \citenamefont
  {Jauho}}]{Furst2009PRB}%
  \BibitemOpen
  \bibfield  {author} {\bibinfo {author} {\bibfnamefont {J.}~\bibnamefont
  {F\"{u}rst}}, \bibinfo {author} {\bibfnamefont {T.}~\bibnamefont {Pedersen}},
  \bibinfo {author} {\bibfnamefont {M.}~\bibnamefont {Brandbyge}}, \ and\
  \bibinfo {author} {\bibfnamefont {A.-P.}\ \bibnamefont {Jauho}},\ }\href
  {\doibase 10.1103/PhysRevB.80.115117} {\bibfield  {journal} {\bibinfo
  {journal} {Phys. Rev. B}\ }\textbf {\bibinfo {volume} {80}},\ \bibinfo
  {pages} {115117} (\bibinfo {year} {2009}{\natexlab{b}})}\BibitemShut
  {NoStop}%
\bibitem [{\citenamefont {Vanevi\'{c}}\ \emph {et~al.}(2009)\citenamefont
  {Vanevi\'{c}}, \citenamefont {Stojanovi\'{c}},\ and\ \citenamefont
  {Kindermann}}]{Vanevic2009}%
  \BibitemOpen
  \bibfield  {author} {\bibinfo {author} {\bibfnamefont {M.}~\bibnamefont
  {Vanevi\'{c}}}, \bibinfo {author} {\bibfnamefont {V.}~\bibnamefont
  {Stojanovi\'{c}}}, \ and\ \bibinfo {author} {\bibfnamefont {M.}~\bibnamefont
  {Kindermann}},\ }\href {\doibase 10.1103/PhysRevB.80.045410} {\bibfield
  {journal} {\bibinfo  {journal} {Phys. Rev. B}\ }\textbf {\bibinfo {volume}
  {80}},\ \bibinfo {pages} {045410} (\bibinfo {year} {2009})}\BibitemShut
  {NoStop}%
\bibitem [{\citenamefont {Rosales}\ \emph {et~al.}(2009)\citenamefont
  {Rosales}, \citenamefont {Pacheco}, \citenamefont {Barticevic}, \citenamefont
  {Le\'{o}n}, \citenamefont {Latg\'{e}},\ and\ \citenamefont
  {Orellana}}]{Rosales2009}%
  \BibitemOpen
  \bibfield  {author} {\bibinfo {author} {\bibfnamefont {L.}~\bibnamefont
  {Rosales}}, \bibinfo {author} {\bibfnamefont {M.}~\bibnamefont {Pacheco}},
  \bibinfo {author} {\bibfnamefont {Z.}~\bibnamefont {Barticevic}}, \bibinfo
  {author} {\bibfnamefont {A.}~\bibnamefont {Le\'{o}n}}, \bibinfo {author}
  {\bibfnamefont {A.}~\bibnamefont {Latg\'{e}}}, \ and\ \bibinfo {author}
  {\bibfnamefont {P.}~\bibnamefont {Orellana}},\ }\href {\doibase
  10.1103/PhysRevB.80.073402} {\bibfield  {journal} {\bibinfo  {journal} {Phys.
  Rev. B}\ }\textbf {\bibinfo {volume} {80}},\ \bibinfo {pages} {073402}
  (\bibinfo {year} {2009})}\BibitemShut {NoStop}%
\bibitem [{\citenamefont {Zheng}\ \emph {et~al.}(2009)\citenamefont {Zheng},
  \citenamefont {Zhang}, \citenamefont {Zeng}, \citenamefont
  {Garc\'{\i}a-Su\'{a}rez},\ and\ \citenamefont {Lambert}}]{Zheng2009}%
  \BibitemOpen
  \bibfield  {author} {\bibinfo {author} {\bibfnamefont {X.}~\bibnamefont
  {Zheng}}, \bibinfo {author} {\bibfnamefont {G.}~\bibnamefont {Zhang}},
  \bibinfo {author} {\bibfnamefont {Z.}~\bibnamefont {Zeng}}, \bibinfo {author}
  {\bibfnamefont {V.}~\bibnamefont {Garc\'{\i}a-Su\'{a}rez}}, \ and\ \bibinfo
  {author} {\bibfnamefont {C.}~\bibnamefont {Lambert}},\ }\href {\doibase
  10.1103/PhysRevB.80.075413} {\bibfield  {journal} {\bibinfo  {journal} {Phys.
  Rev. B}\ }\textbf {\bibinfo {volume} {80}},\ \bibinfo {pages} {075413}
  (\bibinfo {year} {2009})}\BibitemShut {NoStop}%
\bibitem [{\citenamefont {Ritter}\ \emph {et~al.}(2009)\citenamefont {Ritter},
  \citenamefont {Pacheco}, \citenamefont {Orellana},\ and\ \citenamefont
  {Latg\'{e}}}]{Ritter2009}%
  \BibitemOpen
  \bibfield  {author} {\bibinfo {author} {\bibfnamefont {C.}~\bibnamefont
  {Ritter}}, \bibinfo {author} {\bibfnamefont {M.}~\bibnamefont {Pacheco}},
  \bibinfo {author} {\bibfnamefont {P.}~\bibnamefont {Orellana}}, \ and\
  \bibinfo {author} {\bibfnamefont {A.}~\bibnamefont {Latg\'{e}}},\ }\href
  {\doibase 10.1063/1.3259408} {\bibfield  {journal} {\bibinfo  {journal}
  {Journal of Applied Physics}\ }\textbf {\bibinfo {volume} {106}},\ \bibinfo
  {pages} {104303} (\bibinfo {year} {2009})}\BibitemShut {NoStop}%
\bibitem [{\citenamefont {Petersen}\ \emph {et~al.}(2011)\citenamefont
  {Petersen}, \citenamefont {Pedersen},\ and\ \citenamefont
  {Jauho}}]{ClarSextetGAL}%
  \BibitemOpen
  \bibfield  {author} {\bibinfo {author} {\bibfnamefont {R.}~\bibnamefont
  {Petersen}}, \bibinfo {author} {\bibfnamefont {T.~G.}\ \bibnamefont
  {Pedersen}}, \ and\ \bibinfo {author} {\bibfnamefont {A.-P.}\ \bibnamefont
  {Jauho}},\ }\href {\doibase 10.1021/nn102442h} {\bibfield  {journal}
  {\bibinfo  {journal} {ACS Nano}\ }\textbf {\bibinfo {volume} {5}},\ \bibinfo
  {pages} {523} (\bibinfo {year} {2011})}\BibitemShut {NoStop}%
\bibitem [{\citenamefont {Ouyang}\ \emph {et~al.}(2011)\citenamefont {Ouyang},
  \citenamefont {Peng}, \citenamefont {Liu},\ and\ \citenamefont
  {Liu}}]{Ouyang2011}%
  \BibitemOpen
  \bibfield  {author} {\bibinfo {author} {\bibfnamefont {F.}~\bibnamefont
  {Ouyang}}, \bibinfo {author} {\bibfnamefont {S.}~\bibnamefont {Peng}},
  \bibinfo {author} {\bibfnamefont {Z.}~\bibnamefont {Liu}}, \ and\ \bibinfo
  {author} {\bibfnamefont {Z.}~\bibnamefont {Liu}},\ }\href {\doibase
  10.1021/nn200580w} {\bibfield  {journal} {\bibinfo  {journal} {ACS Nano}\
  }\textbf {\bibinfo {volume} {5}},\ \bibinfo {pages} {4023} (\bibinfo {year}
  {2011})}\BibitemShut {NoStop}%
\bibitem [{\citenamefont {Gunst}\ \emph {et~al.}(2011)\citenamefont {Gunst},
  \citenamefont {Markussen}, \citenamefont {Jauho},\ and\ \citenamefont
  {Brandbyge}}]{Tue:finiteGAL}%
  \BibitemOpen
  \bibfield  {author} {\bibinfo {author} {\bibfnamefont {T.}~\bibnamefont
  {Gunst}}, \bibinfo {author} {\bibfnamefont {T.}~\bibnamefont {Markussen}},
  \bibinfo {author} {\bibfnamefont {A.-P.}\ \bibnamefont {Jauho}}, \ and\
  \bibinfo {author} {\bibfnamefont {M.}~\bibnamefont {Brandbyge}},\ }\href
  {\doibase 10.1103/PhysRevB.84.155449} {\bibfield  {journal} {\bibinfo
  {journal} {Phys. Rev. B}\ }\textbf {\bibinfo {volume} {84}},\ \bibinfo
  {pages} {155449} (\bibinfo {year} {2011})}\BibitemShut {NoStop}%
\bibitem [{\citenamefont {Pedersen}\ and\ \citenamefont
  {Pedersen}(2012{\natexlab{a}})}]{Pedersen2012a}%
  \BibitemOpen
  \bibfield  {author} {\bibinfo {author} {\bibfnamefont {T.~G.}\ \bibnamefont
  {Pedersen}}\ and\ \bibinfo {author} {\bibfnamefont {J.~G.}\ \bibnamefont
  {Pedersen}},\ }\href {\doibase 10.1063/1.4768844} {\bibfield  {journal}
  {\bibinfo  {journal} {Journal of Applied Physics}\ }\textbf {\bibinfo
  {volume} {112}},\ \bibinfo {pages} {113715} (\bibinfo {year}
  {2012}{\natexlab{a}})}\BibitemShut {NoStop}%
\bibitem [{\citenamefont {Pedersen}\ \emph {et~al.}(2012)\citenamefont
  {Pedersen}, \citenamefont {Gunst}, \citenamefont {Markussen},\ and\
  \citenamefont {Pedersen}}]{GALWG}%
  \BibitemOpen
  \bibfield  {author} {\bibinfo {author} {\bibfnamefont {J.~G.}\ \bibnamefont
  {Pedersen}}, \bibinfo {author} {\bibfnamefont {T.}~\bibnamefont {Gunst}},
  \bibinfo {author} {\bibfnamefont {T.}~\bibnamefont {Markussen}}, \ and\
  \bibinfo {author} {\bibfnamefont {T.~G.}\ \bibnamefont {Pedersen}},\ }\href
  {\doibase 10.1103/PhysRevB.86.245410} {\bibfield  {journal} {\bibinfo
  {journal} {Phys. Rev. B}\ }\textbf {\bibinfo {volume} {86}},\ \bibinfo
  {pages} {245410} (\bibinfo {year} {2012})}\BibitemShut {NoStop}%
\bibitem [{\citenamefont {Chang}\ and\ \citenamefont
  {Nikoli\'{c}}(2012)}]{Chang2012}%
  \BibitemOpen
  \bibfield  {author} {\bibinfo {author} {\bibfnamefont {P.-H.}\ \bibnamefont
  {Chang}}\ and\ \bibinfo {author} {\bibfnamefont {B.~K.}\ \bibnamefont
  {Nikoli\'{c}}},\ }\href {\doibase 10.1103/PhysRevB.86.041406} {\bibfield
  {journal} {\bibinfo  {journal} {Phys. Rev. B}\ }\textbf {\bibinfo {volume}
  {86}},\ \bibinfo {pages} {041406} (\bibinfo {year} {2012})}\BibitemShut
  {NoStop}%
\bibitem [{\citenamefont {Yuan}\ \emph
  {et~al.}(2013{\natexlab{a}})\citenamefont {Yuan}, \citenamefont {Rold\'{a}n},
  \citenamefont {Jauho},\ and\ \citenamefont {Katsnelson}}]{Yuan_GAL_disorder}%
  \BibitemOpen
  \bibfield  {author} {\bibinfo {author} {\bibfnamefont {S.}~\bibnamefont
  {Yuan}}, \bibinfo {author} {\bibfnamefont {R.}~\bibnamefont {Rold\'{a}n}},
  \bibinfo {author} {\bibfnamefont {A.-P.}\ \bibnamefont {Jauho}}, \ and\
  \bibinfo {author} {\bibfnamefont {M.}~\bibnamefont {Katsnelson}},\ }\href
  {\doibase 10.1103/PhysRevB.87.085430} {\bibfield  {journal} {\bibinfo
  {journal} {Phys. Rev. B}\ }\textbf {\bibinfo {volume} {87}},\ \bibinfo
  {pages} {085430} (\bibinfo {year} {2013}{\natexlab{a}})}\BibitemShut
  {NoStop}%
\bibitem [{\citenamefont {Yuan}\ \emph
  {et~al.}(2013{\natexlab{b}})\citenamefont {Yuan}, \citenamefont {Jin},
  \citenamefont {Rold\'{a}n}, \citenamefont {Jauho},\ and\ \citenamefont
  {Katsnelson}}]{Yuan_GAL_screening}%
  \BibitemOpen
  \bibfield  {author} {\bibinfo {author} {\bibfnamefont {S.}~\bibnamefont
  {Yuan}}, \bibinfo {author} {\bibfnamefont {F.}~\bibnamefont {Jin}}, \bibinfo
  {author} {\bibfnamefont {R.}~\bibnamefont {Rold\'{a}n}}, \bibinfo {author}
  {\bibfnamefont {A.-P.}\ \bibnamefont {Jauho}}, \ and\ \bibinfo {author}
  {\bibfnamefont {M.~I.}\ \bibnamefont {Katsnelson}},\ }\href {\doibase
  10.1103/PhysRevB.88.195401} {\bibfield  {journal} {\bibinfo  {journal} {Phys.
  Rev. B}\ }\textbf {\bibinfo {volume} {88}},\ \bibinfo {pages} {195401}
  (\bibinfo {year} {2013}{\natexlab{b}})}\BibitemShut {NoStop}%
\bibitem [{\citenamefont {Liu}\ \emph {et~al.}(2013)\citenamefont {Liu},
  \citenamefont {Zhang},\ and\ \citenamefont {Guo}}]{Liu2013}%
  \BibitemOpen
  \bibfield  {author} {\bibinfo {author} {\bibfnamefont {X.}~\bibnamefont
  {Liu}}, \bibinfo {author} {\bibfnamefont {Z.}~\bibnamefont {Zhang}}, \ and\
  \bibinfo {author} {\bibfnamefont {W.}~\bibnamefont {Guo}},\ }\href {\doibase
  10.1002/smll.201202988} {\bibfield  {journal} {\bibinfo  {journal} {Small}\
  }\textbf {\bibinfo {volume} {9}},\ \bibinfo {pages} {1405} (\bibinfo {year}
  {2013})}\BibitemShut {NoStop}%
\bibitem [{\citenamefont {{Hung Nguyen}}\ \emph {et~al.}(2013)\citenamefont
  {{Hung Nguyen}}, \citenamefont {{Chung Nguyen}}, \citenamefont {Nguyen},\
  and\ \citenamefont {Dollfus}}]{HungNguyen2013}%
  \BibitemOpen
  \bibfield  {author} {\bibinfo {author} {\bibfnamefont {V.}~\bibnamefont
  {{Hung Nguyen}}}, \bibinfo {author} {\bibfnamefont {M.}~\bibnamefont {{Chung
  Nguyen}}}, \bibinfo {author} {\bibfnamefont {H.-V.}\ \bibnamefont {Nguyen}},
  \ and\ \bibinfo {author} {\bibfnamefont {P.}~\bibnamefont {Dollfus}},\ }\href
  {\doibase 10.1063/1.4772609} {\bibfield  {journal} {\bibinfo  {journal}
  {Journal of Applied Physics}\ }\textbf {\bibinfo {volume} {113}},\ \bibinfo
  {pages} {013702} (\bibinfo {year} {2013})}\BibitemShut {NoStop}%
\bibitem [{\citenamefont {Trolle}\ \emph {et~al.}(2013)\citenamefont {Trolle},
  \citenamefont {M{\o}ller},\ and\ \citenamefont {Pedersen}}]{Trolle-spinGAL}%
  \BibitemOpen
  \bibfield  {author} {\bibinfo {author} {\bibfnamefont {M.~L.}\ \bibnamefont
  {Trolle}}, \bibinfo {author} {\bibfnamefont {U.~S.}\ \bibnamefont
  {M{\o}ller}}, \ and\ \bibinfo {author} {\bibfnamefont {T.~G.}\ \bibnamefont
  {Pedersen}},\ }\href {\doibase 10.1103/PhysRevB.88.195418} {\bibfield
  {journal} {\bibinfo  {journal} {Phys. Rev. B}\ }\textbf {\bibinfo {volume}
  {88}},\ \bibinfo {pages} {195418} (\bibinfo {year} {2013})}\BibitemShut
  {NoStop}%
\bibitem [{\citenamefont {Dvorak}\ \emph {et~al.}(2013)\citenamefont {Dvorak},
  \citenamefont {Oswald},\ and\ \citenamefont {Wu}}]{Dvorak2013}%
  \BibitemOpen
  \bibfield  {author} {\bibinfo {author} {\bibfnamefont {M.}~\bibnamefont
  {Dvorak}}, \bibinfo {author} {\bibfnamefont {W.}~\bibnamefont {Oswald}}, \
  and\ \bibinfo {author} {\bibfnamefont {Z.}~\bibnamefont {Wu}},\ }\href
  {\doibase 10.1038/srep02289} {\bibfield  {journal} {\bibinfo  {journal}
  {Scientific Reports}\ }\textbf {\bibinfo {volume} {3}},\ \bibinfo {pages}
  {2289} (\bibinfo {year} {2013})}\BibitemShut {NoStop}%
\bibitem [{\citenamefont {Ji}\ \emph {et~al.}(2013)\citenamefont {Ji},
  \citenamefont {Zhang}, \citenamefont {Wang}, \citenamefont {Qian},\ and\
  \citenamefont {Yu}}]{Ji2013}%
  \BibitemOpen
  \bibfield  {author} {\bibinfo {author} {\bibfnamefont {X.}~\bibnamefont
  {Ji}}, \bibinfo {author} {\bibfnamefont {J.}~\bibnamefont {Zhang}}, \bibinfo
  {author} {\bibfnamefont {Y.}~\bibnamefont {Wang}}, \bibinfo {author}
  {\bibfnamefont {H.}~\bibnamefont {Qian}}, \ and\ \bibinfo {author}
  {\bibfnamefont {Z.}~\bibnamefont {Yu}},\ }\href {\doibase 10.1039/c3nr33241a}
  {\bibfield  {journal} {\bibinfo  {journal} {Nanoscale}\ }\textbf {\bibinfo
  {volume} {5}},\ \bibinfo {pages} {2527} (\bibinfo {year} {2013})}\BibitemShut
  {NoStop}%
\bibitem [{\citenamefont {Brun}\ \emph {et~al.}(2014)\citenamefont {Brun},
  \citenamefont {Thomsen},\ and\ \citenamefont {Pedersen}}]{Brun2014}%
  \BibitemOpen
  \bibfield  {author} {\bibinfo {author} {\bibfnamefont {S.~J.}\ \bibnamefont
  {Brun}}, \bibinfo {author} {\bibfnamefont {M.~R.}\ \bibnamefont {Thomsen}}, \
  and\ \bibinfo {author} {\bibfnamefont {T.~G.}\ \bibnamefont {Pedersen}},\
  }\href {\doibase 10.1088/0953-8984/26/26/265301} {\bibfield  {journal}
  {\bibinfo  {journal} {Journal of Physics: Condensed matter}\ }\textbf
  {\bibinfo {volume} {26}},\ \bibinfo {pages} {265301} (\bibinfo {year}
  {2014})}\BibitemShut {NoStop}%
\bibitem [{\citenamefont {Bieri}\ \emph {et~al.}(2009)\citenamefont {Bieri},
  \citenamefont {Treier}, \citenamefont {Cai}, \citenamefont
  {A\"{\i}t-Mansour}, \citenamefont {Ruffieux}, \citenamefont {Gr\"{o}ning},
  \citenamefont {Gr\"{o}ning}, \citenamefont {Kastler}, \citenamefont {Rieger},
  \citenamefont {Feng}, \citenamefont {M\"{u}llen},\ and\ \citenamefont
  {Fasel}}]{Bieri2009}%
  \BibitemOpen
  \bibfield  {author} {\bibinfo {author} {\bibfnamefont {M.}~\bibnamefont
  {Bieri}}, \bibinfo {author} {\bibfnamefont {M.}~\bibnamefont {Treier}},
  \bibinfo {author} {\bibfnamefont {J.}~\bibnamefont {Cai}}, \bibinfo {author}
  {\bibfnamefont {K.}~\bibnamefont {A\"{\i}t-Mansour}}, \bibinfo {author}
  {\bibfnamefont {P.}~\bibnamefont {Ruffieux}}, \bibinfo {author}
  {\bibfnamefont {O.}~\bibnamefont {Gr\"{o}ning}}, \bibinfo {author}
  {\bibfnamefont {P.}~\bibnamefont {Gr\"{o}ning}}, \bibinfo {author}
  {\bibfnamefont {M.}~\bibnamefont {Kastler}}, \bibinfo {author} {\bibfnamefont
  {R.}~\bibnamefont {Rieger}}, \bibinfo {author} {\bibfnamefont
  {X.}~\bibnamefont {Feng}}, \bibinfo {author} {\bibfnamefont {K.}~\bibnamefont
  {M\"{u}llen}}, \ and\ \bibinfo {author} {\bibfnamefont {R.}~\bibnamefont
  {Fasel}},\ }\href {\doibase 10.1039/b915190g} {\bibfield  {journal} {\bibinfo
   {journal} {Chemical communications}\ ,\ \bibinfo {pages} {6919}} (\bibinfo
  {year} {2009})}\BibitemShut {NoStop}%
\bibitem [{\citenamefont {Kim}\ \emph {et~al.}(2010)\citenamefont {Kim},
  \citenamefont {Safron}, \citenamefont {Han}, \citenamefont {Arnold},\ and\
  \citenamefont {Gopalan}}]{Kim2010}%
  \BibitemOpen
  \bibfield  {author} {\bibinfo {author} {\bibfnamefont {M.}~\bibnamefont
  {Kim}}, \bibinfo {author} {\bibfnamefont {N.~S.}\ \bibnamefont {Safron}},
  \bibinfo {author} {\bibfnamefont {E.}~\bibnamefont {Han}}, \bibinfo {author}
  {\bibfnamefont {M.~S.}\ \bibnamefont {Arnold}}, \ and\ \bibinfo {author}
  {\bibfnamefont {P.}~\bibnamefont {Gopalan}},\ }\href {\doibase
  10.1021/nl9032318} {\bibfield  {journal} {\bibinfo  {journal} {Nano Letters}\
  }\textbf {\bibinfo {volume} {10}},\ \bibinfo {pages} {1125} (\bibinfo {year}
  {2010})}\BibitemShut {NoStop}%
\bibitem [{\citenamefont {Bai}\ \emph {et~al.}(2010)\citenamefont {Bai},
  \citenamefont {Zhong}, \citenamefont {Jiang}, \citenamefont {Huang},\ and\
  \citenamefont {Duan}}]{Bai2010}%
  \BibitemOpen
  \bibfield  {author} {\bibinfo {author} {\bibfnamefont {J.}~\bibnamefont
  {Bai}}, \bibinfo {author} {\bibfnamefont {X.}~\bibnamefont {Zhong}}, \bibinfo
  {author} {\bibfnamefont {S.}~\bibnamefont {Jiang}}, \bibinfo {author}
  {\bibfnamefont {Y.}~\bibnamefont {Huang}}, \ and\ \bibinfo {author}
  {\bibfnamefont {X.}~\bibnamefont {Duan}},\ }\href {\doibase
  10.1038/nnano.2010.8} {\bibfield  {journal} {\bibinfo  {journal} {Nature
  Nanotechnology}\ }\textbf {\bibinfo {volume} {5}},\ \bibinfo {pages} {190}
  (\bibinfo {year} {2010})}\BibitemShut {NoStop}%
\bibitem [{\citenamefont {Kim}\ \emph {et~al.}(2012)\citenamefont {Kim},
  \citenamefont {Safron}, \citenamefont {Han}, \citenamefont {Arnold},\ and\
  \citenamefont {Gopalan}}]{Kim2012}%
  \BibitemOpen
  \bibfield  {author} {\bibinfo {author} {\bibfnamefont {M.}~\bibnamefont
  {Kim}}, \bibinfo {author} {\bibfnamefont {N.~S.}\ \bibnamefont {Safron}},
  \bibinfo {author} {\bibfnamefont {E.}~\bibnamefont {Han}}, \bibinfo {author}
  {\bibfnamefont {M.~S.}\ \bibnamefont {Arnold}}, \ and\ \bibinfo {author}
  {\bibfnamefont {P.}~\bibnamefont {Gopalan}},\ }\href {\doibase
  10.1021/nn3033985} {\bibfield  {journal} {\bibinfo  {journal} {ACS Nano}\
  }\textbf {\bibinfo {volume} {6}},\ \bibinfo {pages} {9846} (\bibinfo {year}
  {2012})}\BibitemShut {NoStop}%
\bibitem [{\citenamefont {Shen}\ \emph {et~al.}(2008)\citenamefont {Shen},
  \citenamefont {Wu}, \citenamefont {Capano}, \citenamefont {Rokhinson},
  \citenamefont {Engel},\ and\ \citenamefont {Ye}}]{Shen2008}%
  \BibitemOpen
  \bibfield  {author} {\bibinfo {author} {\bibfnamefont {T.}~\bibnamefont
  {Shen}}, \bibinfo {author} {\bibfnamefont {Y.~Q.}\ \bibnamefont {Wu}},
  \bibinfo {author} {\bibfnamefont {M.~A.}\ \bibnamefont {Capano}}, \bibinfo
  {author} {\bibfnamefont {L.~P.}\ \bibnamefont {Rokhinson}}, \bibinfo {author}
  {\bibfnamefont {L.~W.}\ \bibnamefont {Engel}}, \ and\ \bibinfo {author}
  {\bibfnamefont {P.~D.}\ \bibnamefont {Ye}},\ }\href {\doibase
  10.1063/1.2988725} {\bibfield  {journal} {\bibinfo  {journal} {Applied
  Physics Letters}\ }\textbf {\bibinfo {volume} {93}},\ \bibinfo {pages}
  {122102} (\bibinfo {year} {2008})}\BibitemShut {NoStop}%
\bibitem [{\citenamefont {Eroms}\ and\ \citenamefont
  {Weiss}(2009)}]{Eroms2009}%
  \BibitemOpen
  \bibfield  {author} {\bibinfo {author} {\bibfnamefont {J.}~\bibnamefont
  {Eroms}}\ and\ \bibinfo {author} {\bibfnamefont {D.}~\bibnamefont {Weiss}},\
  }\href {\doibase 10.1088/1367-2630/11/9/095021} {\bibfield  {journal}
  {\bibinfo  {journal} {New Journal of Physics}\ }\textbf {\bibinfo {volume}
  {11}},\ \bibinfo {pages} {095021} (\bibinfo {year} {2009})}\BibitemShut
  {NoStop}%
\bibitem [{\citenamefont {Begliarbekov}\ \emph {et~al.}(2011)\citenamefont
  {Begliarbekov}, \citenamefont {Sul}, \citenamefont {Santanello},
  \citenamefont {Ai}, \citenamefont {Zhang}, \citenamefont {Yang},\ and\
  \citenamefont {Strauf}}]{Begliarbekov2011}%
  \BibitemOpen
  \bibfield  {author} {\bibinfo {author} {\bibfnamefont {M.}~\bibnamefont
  {Begliarbekov}}, \bibinfo {author} {\bibfnamefont {O.}~\bibnamefont {Sul}},
  \bibinfo {author} {\bibfnamefont {J.}~\bibnamefont {Santanello}}, \bibinfo
  {author} {\bibfnamefont {N.}~\bibnamefont {Ai}}, \bibinfo {author}
  {\bibfnamefont {X.}~\bibnamefont {Zhang}}, \bibinfo {author} {\bibfnamefont
  {E.-H.}\ \bibnamefont {Yang}}, \ and\ \bibinfo {author} {\bibfnamefont
  {S.}~\bibnamefont {Strauf}},\ }\href {\doibase 10.1021/nl1042648} {\bibfield
  {journal} {\bibinfo  {journal} {Nano Letters}\ }\textbf {\bibinfo {volume}
  {11}},\ \bibinfo {pages} {1254} (\bibinfo {year} {2011})}\BibitemShut
  {NoStop}%
\bibitem [{\citenamefont {Giesbers}\ \emph {et~al.}(2012)\citenamefont
  {Giesbers}, \citenamefont {Peters}, \citenamefont {Burghard},\ and\
  \citenamefont {Kern}}]{Giesbers2012}%
  \BibitemOpen
  \bibfield  {author} {\bibinfo {author} {\bibfnamefont {A.~J.~M.}\
  \bibnamefont {Giesbers}}, \bibinfo {author} {\bibfnamefont {E.~C.}\
  \bibnamefont {Peters}}, \bibinfo {author} {\bibfnamefont {M.}~\bibnamefont
  {Burghard}}, \ and\ \bibinfo {author} {\bibfnamefont {K.}~\bibnamefont
  {Kern}},\ }\href {\doibase 10.1103/PhysRevB.86.045445} {\bibfield  {journal}
  {\bibinfo  {journal} {Phys. Rev. B}\ }\textbf {\bibinfo {volume} {86}},\
  \bibinfo {pages} {045445} (\bibinfo {year} {2012})}\BibitemShut {NoStop}%
\bibitem [{\citenamefont {Oberhuber}\ \emph {et~al.}(2013)\citenamefont
  {Oberhuber}, \citenamefont {Blien}, \citenamefont {Heydrich}, \citenamefont
  {Yaghobian}, \citenamefont {Korn}, \citenamefont {Schüller}, \citenamefont
  {Strunk}, \citenamefont {Weiss},\ and\ \citenamefont
  {Eroms}}]{Oberhuber2013}%
  \BibitemOpen
  \bibfield  {author} {\bibinfo {author} {\bibfnamefont {F.}~\bibnamefont
  {Oberhuber}}, \bibinfo {author} {\bibfnamefont {S.}~\bibnamefont {Blien}},
  \bibinfo {author} {\bibfnamefont {S.}~\bibnamefont {Heydrich}}, \bibinfo
  {author} {\bibfnamefont {F.}~\bibnamefont {Yaghobian}}, \bibinfo {author}
  {\bibfnamefont {T.}~\bibnamefont {Korn}}, \bibinfo {author} {\bibfnamefont
  {C.}~\bibnamefont {Schüller}}, \bibinfo {author} {\bibfnamefont
  {C.}~\bibnamefont {Strunk}}, \bibinfo {author} {\bibfnamefont
  {D.}~\bibnamefont {Weiss}}, \ and\ \bibinfo {author} {\bibfnamefont
  {J.}~\bibnamefont {Eroms}},\ }\href {\doibase 10.1063/1.4824025} {\bibfield
  {journal} {\bibinfo  {journal} {Applied Physics Letters}\ }\textbf {\bibinfo
  {volume} {103}},\ \bibinfo {pages} {143111} (\bibinfo {year}
  {2013})}\BibitemShut {NoStop}%
\bibitem [{\citenamefont {Xu}\ \emph {et~al.}(2013)\citenamefont {Xu},
  \citenamefont {Wu}, \citenamefont {Schneider}, \citenamefont {Houben},
  \citenamefont {Malladi}, \citenamefont {Dekker}, \citenamefont {Yucelen},
  \citenamefont {Dunin-Borkowski},\ and\ \citenamefont {Zandbergen}}]{Xu2013}%
  \BibitemOpen
  \bibfield  {author} {\bibinfo {author} {\bibfnamefont {Q.}~\bibnamefont
  {Xu}}, \bibinfo {author} {\bibfnamefont {M.-Y.}\ \bibnamefont {Wu}}, \bibinfo
  {author} {\bibfnamefont {G.~F.}\ \bibnamefont {Schneider}}, \bibinfo {author}
  {\bibfnamefont {L.}~\bibnamefont {Houben}}, \bibinfo {author} {\bibfnamefont
  {S.~K.}\ \bibnamefont {Malladi}}, \bibinfo {author} {\bibfnamefont
  {C.}~\bibnamefont {Dekker}}, \bibinfo {author} {\bibfnamefont
  {E.}~\bibnamefont {Yucelen}}, \bibinfo {author} {\bibfnamefont {R.~E.}\
  \bibnamefont {Dunin-Borkowski}}, \ and\ \bibinfo {author} {\bibfnamefont
  {H.~W.}\ \bibnamefont {Zandbergen}},\ }\href {\doibase 10.1021/nn3053582}
  {\bibfield  {journal} {\bibinfo  {journal} {ACS Nano}\ }\textbf {\bibinfo
  {volume} {7}},\ \bibinfo {pages} {1566} (\bibinfo {year} {2013})}\BibitemShut
  {NoStop}%
\bibitem [{\citenamefont {Low}\ \emph {et~al.}(2011)\citenamefont {Low},
  \citenamefont {Guinea},\ and\ \citenamefont {Katsnelson}}]{Low2011}%
  \BibitemOpen
  \bibfield  {author} {\bibinfo {author} {\bibfnamefont {T.}~\bibnamefont
  {Low}}, \bibinfo {author} {\bibfnamefont {F.}~\bibnamefont {Guinea}}, \ and\
  \bibinfo {author} {\bibfnamefont {M.~I.}\ \bibnamefont {Katsnelson}},\ }\href
  {\doibase 10.1103/PhysRevB.83.195436} {\bibfield  {journal} {\bibinfo
  {journal} {Phys. Rev. B}\ }\textbf {\bibinfo {volume} {83}},\ \bibinfo
  {pages} {195436} (\bibinfo {year} {2011})}\BibitemShut {NoStop}%
\bibitem [{\citenamefont {Pedersen}\ and\ \citenamefont
  {Pedersen}(2012{\natexlab{b}})}]{Pedersen2012}%
  \BibitemOpen
  \bibfield  {author} {\bibinfo {author} {\bibfnamefont {J.~G.}\ \bibnamefont
  {Pedersen}}\ and\ \bibinfo {author} {\bibfnamefont {T.~G.}\ \bibnamefont
  {Pedersen}},\ }\href {\doibase 10.1103/PhysRevB.85.235432} {\bibfield
  {journal} {\bibinfo  {journal} {Phys. Rev. B}\ }\textbf {\bibinfo {volume}
  {85}},\ \bibinfo {pages} {235432} (\bibinfo {year}
  {2012}{\natexlab{b}})}\BibitemShut {NoStop}%
\bibitem [{\citenamefont {Pereira}\ and\ \citenamefont
  {Castro~Neto}(2009)}]{pereira_strain_2009}%
  \BibitemOpen
  \bibfield  {author} {\bibinfo {author} {\bibfnamefont {V.~M.}\ \bibnamefont
  {Pereira}}\ and\ \bibinfo {author} {\bibfnamefont {A.~H.}\ \bibnamefont
  {Castro~Neto}},\ }\href {\doibase 10.1103/PhysRevLett.103.046801} {\bibfield
  {journal} {\bibinfo  {journal} {Phys. Rev. Lett.}\ }\textbf {\bibinfo
  {volume} {103}},\ \bibinfo {pages} {046801} (\bibinfo {year}
  {2009})}\BibitemShut {NoStop}%
\bibitem [{\citenamefont {Pereira}\ \emph {et~al.}(2009)\citenamefont
  {Pereira}, \citenamefont {Castro~Neto},\ and\ \citenamefont
  {Peres}}]{pereira_tight-binding_2009}%
  \BibitemOpen
  \bibfield  {author} {\bibinfo {author} {\bibfnamefont {V.~M.}\ \bibnamefont
  {Pereira}}, \bibinfo {author} {\bibfnamefont {A.~H.}\ \bibnamefont
  {Castro~Neto}}, \ and\ \bibinfo {author} {\bibfnamefont {N.~M.~R.}\
  \bibnamefont {Peres}},\ }\href {\doibase 10.1103/PhysRevB.80.045401}
  {\bibfield  {journal} {\bibinfo  {journal} {Phys. Rev. B}\ }\textbf {\bibinfo
  {volume} {80}},\ \bibinfo {pages} {045401} (\bibinfo {year}
  {2009})}\BibitemShut {NoStop}%
\bibitem [{\citenamefont {Katsnelson}\ \emph {et~al.}(2006)\citenamefont
  {Katsnelson}, \citenamefont {Novoselov},\ and\ \citenamefont
  {Geim}}]{katsnelson_chiral_2006}%
  \BibitemOpen
  \bibfield  {author} {\bibinfo {author} {\bibfnamefont {M.~I.}\ \bibnamefont
  {Katsnelson}}, \bibinfo {author} {\bibfnamefont {K.~S.}\ \bibnamefont
  {Novoselov}}, \ and\ \bibinfo {author} {\bibfnamefont {A.~K.}\ \bibnamefont
  {Geim}},\ }\href {\doibase 10.1038/nphys384} {\bibfield  {journal} {\bibinfo
  {journal} {Nature Physics}\ }\textbf {\bibinfo {volume} {2}},\ \bibinfo
  {pages} {620} (\bibinfo {year} {2006})}\BibitemShut {NoStop}%
\bibitem [{\citenamefont {Chutinan}\ and\ \citenamefont
  {Noda}(2000)}]{Chutinan2000}%
  \BibitemOpen
  \bibfield  {author} {\bibinfo {author} {\bibfnamefont {A.}~\bibnamefont
  {Chutinan}}\ and\ \bibinfo {author} {\bibfnamefont {S.}~\bibnamefont
  {Noda}},\ }\href {\doibase 10.1103/PhysRevB.62.4488} {\bibfield  {journal}
  {\bibinfo  {journal} {Phys. Rev. B}\ }\textbf {\bibinfo {volume} {62}},\
  \bibinfo {pages} {4488} (\bibinfo {year} {2000})}\BibitemShut {NoStop}%
\bibitem [{\citenamefont {Jia}\ \emph {et~al.}(2009)\citenamefont {Jia},
  \citenamefont {Hofmann}, \citenamefont {Meunier}, \citenamefont {Sumpter},
  \citenamefont {Campos-Delgado}, \citenamefont {Romo-Herrera}, \citenamefont
  {Son}, \citenamefont {Hsieh}, \citenamefont {Reina}, \citenamefont {Kong},
  \citenamefont {Terrones},\ and\ \citenamefont {Dresselhaus}}]{Jia2009}%
  \BibitemOpen
  \bibfield  {author} {\bibinfo {author} {\bibfnamefont {X.}~\bibnamefont
  {Jia}}, \bibinfo {author} {\bibfnamefont {M.}~\bibnamefont {Hofmann}},
  \bibinfo {author} {\bibfnamefont {V.}~\bibnamefont {Meunier}}, \bibinfo
  {author} {\bibfnamefont {B.~G.}\ \bibnamefont {Sumpter}}, \bibinfo {author}
  {\bibfnamefont {J.}~\bibnamefont {Campos-Delgado}}, \bibinfo {author}
  {\bibfnamefont {J.~M.}\ \bibnamefont {Romo-Herrera}}, \bibinfo {author}
  {\bibfnamefont {H.}~\bibnamefont {Son}}, \bibinfo {author} {\bibfnamefont
  {Y.-P.}\ \bibnamefont {Hsieh}}, \bibinfo {author} {\bibfnamefont
  {A.}~\bibnamefont {Reina}}, \bibinfo {author} {\bibfnamefont
  {J.}~\bibnamefont {Kong}}, \bibinfo {author} {\bibfnamefont {M.}~\bibnamefont
  {Terrones}}, \ and\ \bibinfo {author} {\bibfnamefont {M.~S.}\ \bibnamefont
  {Dresselhaus}},\ }\href {\doibase 10.1126/science.1166862} {\bibfield
  {journal} {\bibinfo  {journal} {Science}\ }\textbf {\bibinfo {volume}
  {323}},\ \bibinfo {pages} {1701} (\bibinfo {year} {2009})}\BibitemShut
  {NoStop}%
\bibitem [{\citenamefont {Pizzocchero}\ \emph {et~al.}(2014)\citenamefont
  {Pizzocchero}, \citenamefont {Vanin}, \citenamefont {Kling}, \citenamefont
  {Hansen}, \citenamefont {Jacobsen}, \citenamefont {B{\o}ggild},\ and\
  \citenamefont {Booth}}]{Pizzocchero2014}%
  \BibitemOpen
  \bibfield  {author} {\bibinfo {author} {\bibfnamefont {F.}~\bibnamefont
  {Pizzocchero}}, \bibinfo {author} {\bibfnamefont {M.}~\bibnamefont {Vanin}},
  \bibinfo {author} {\bibfnamefont {J.}~\bibnamefont {Kling}}, \bibinfo
  {author} {\bibfnamefont {T.~W.}\ \bibnamefont {Hansen}}, \bibinfo {author}
  {\bibfnamefont {K.~W.}\ \bibnamefont {Jacobsen}}, \bibinfo {author}
  {\bibfnamefont {P.}~\bibnamefont {B{\o}ggild}}, \ and\ \bibinfo {author}
  {\bibfnamefont {T.~J.}\ \bibnamefont {Booth}},\ }\href {\doibase
  10.1021/jp500800n} {\bibfield  {journal} {\bibinfo  {journal} {The Journal of
  Physical Chemistry C}\ }\textbf {\bibinfo {volume} {118}},\ \bibinfo {pages}
  {4296} (\bibinfo {year} {2014})}\BibitemShut {NoStop}%
\bibitem [{\citenamefont {Mucciolo}\ \emph {et~al.}(2009)\citenamefont
  {Mucciolo}, \citenamefont {Castro~Neto},\ and\ \citenamefont
  {Lewenkopf}}]{mucciolo:graphenetransportgaps}%
  \BibitemOpen
  \bibfield  {author} {\bibinfo {author} {\bibfnamefont {E.~R.}\ \bibnamefont
  {Mucciolo}}, \bibinfo {author} {\bibfnamefont {A.~H.}\ \bibnamefont
  {Castro~Neto}}, \ and\ \bibinfo {author} {\bibfnamefont {C.~H.}\ \bibnamefont
  {Lewenkopf}},\ }\href@noop {} {\bibfield  {journal} {\bibinfo  {journal}
  {Phys. Rev. B}\ }\textbf {\bibinfo {volume} {79}},\ \bibinfo {eid} {075407}
  (\bibinfo {year} {2009})}\BibitemShut {NoStop}%
\bibitem [{\citenamefont {Evaldsson}\ \emph {et~al.}(2008)\citenamefont
  {Evaldsson}, \citenamefont {Zozoulenko}, \citenamefont {Xu},\ and\
  \citenamefont {Heinzel}}]{evaldsson:ribbonedgeanderson}%
  \BibitemOpen
  \bibfield  {author} {\bibinfo {author} {\bibfnamefont {M.}~\bibnamefont
  {Evaldsson}}, \bibinfo {author} {\bibfnamefont {I.~V.}\ \bibnamefont
  {Zozoulenko}}, \bibinfo {author} {\bibfnamefont {H.}~\bibnamefont {Xu}}, \
  and\ \bibinfo {author} {\bibfnamefont {T.}~\bibnamefont {Heinzel}},\
  }\href@noop {} {\bibfield  {journal} {\bibinfo  {journal} {Phys. Rev. B}\
  }\textbf {\bibinfo {volume} {78}},\ \bibinfo {eid} {161407} (\bibinfo {year}
  {2008})}\BibitemShut {NoStop}%
\bibitem [{\citenamefont {Sancho}\ \emph {et~al.}(1984)\citenamefont {Sancho},
  \citenamefont {Sancho},\ and\ \citenamefont {Rubio}}]{Sancho-Rubio}%
  \BibitemOpen
  \bibfield  {author} {\bibinfo {author} {\bibfnamefont {M.~P.~L.}\
  \bibnamefont {Sancho}}, \bibinfo {author} {\bibfnamefont {J.~M.~L.}\
  \bibnamefont {Sancho}}, \ and\ \bibinfo {author} {\bibfnamefont
  {J.}~\bibnamefont {Rubio}},\ }\href@noop {} {\bibfield  {journal} {\bibinfo
  {journal} {Journal of Physics F: Metal Physics}\ }\textbf {\bibinfo {volume}
  {14}},\ \bibinfo {pages} {1205} (\bibinfo {year} {1984})}\BibitemShut
  {NoStop}%
\bibitem [{\citenamefont {Lewenkopf}\ and\ \citenamefont
  {Mucciolo}(2013)}]{Lewenkopf2013}%
  \BibitemOpen
  \bibfield  {author} {\bibinfo {author} {\bibfnamefont {C.~H.}\ \bibnamefont
  {Lewenkopf}}\ and\ \bibinfo {author} {\bibfnamefont {E.~R.}\ \bibnamefont
  {Mucciolo}},\ }\href {\doibase 10.1007/s10825-013-0458-7} {\bibfield
  {journal} {\bibinfo  {journal} {Journal of Computational Electronics}\
  }\textbf {\bibinfo {volume} {12}},\ \bibinfo {pages} {203} (\bibinfo {year}
  {2013})}\BibitemShut {NoStop}%
\bibitem [{\citenamefont {Datta}(1997)}]{DattaBook}%
  \BibitemOpen
  \bibfield  {author} {\bibinfo {author} {\bibfnamefont {S.}~\bibnamefont
  {Datta}},\ }\href@noop {} {\emph {\bibinfo {title} {Electronic Transport in
  Mesoscopic Systems}}}\ (\bibinfo  {publisher} {Cambridge University Press},\
  \bibinfo {year} {1997})\BibitemShut {NoStop}%
\bibitem [{per()}]{percolation_footnote}%
  \BibitemOpen
  \href@noop {} {}\bibinfo {note} {Fitting the LDOS map for the pink diamond
  symbol in Figure 3 to a simple site-percolation model gives an occupation
  probability $p \approx 0.77$. This is greater than the critical value, $p_c
  \approx 0.59$ predicted for the onset of percolation in such a
  geometry.}\BibitemShut {Stop}%
\bibitem [{\citenamefont {Wimmer}\ \emph {et~al.}(2010)\citenamefont {Wimmer},
  \citenamefont {Akhmerov},\ and\ \citenamefont {Guinea}}]{Wimmer2010}%
  \BibitemOpen
  \bibfield  {author} {\bibinfo {author} {\bibfnamefont {M.}~\bibnamefont
  {Wimmer}}, \bibinfo {author} {\bibfnamefont {A.~R.}\ \bibnamefont
  {Akhmerov}}, \ and\ \bibinfo {author} {\bibfnamefont {F.}~\bibnamefont
  {Guinea}},\ }\href {\doibase 10.1103/PhysRevB.82.045409} {\bibfield
  {journal} {\bibinfo  {journal} {Phys. Rev. B}\ }\textbf {\bibinfo {volume}
  {82}},\ \bibinfo {pages} {045409} (\bibinfo {year} {2010})}\BibitemShut
  {NoStop}%
\bibitem [{\citenamefont {Kunstmann}\ \emph {et~al.}(2011)\citenamefont
  {Kunstmann}, \citenamefont {\"{O}zdoğan}, \citenamefont {Quandt},\ and\
  \citenamefont {Fehske}}]{Kunstmann2011}%
  \BibitemOpen
  \bibfield  {author} {\bibinfo {author} {\bibfnamefont {J.}~\bibnamefont
  {Kunstmann}}, \bibinfo {author} {\bibfnamefont {C.}~\bibnamefont
  {\"{O}zdoğan}}, \bibinfo {author} {\bibfnamefont {A.}~\bibnamefont
  {Quandt}}, \ and\ \bibinfo {author} {\bibfnamefont {H.}~\bibnamefont
  {Fehske}},\ }\href {\doibase 10.1103/PhysRevB.83.045414} {\bibfield
  {journal} {\bibinfo  {journal} {Phys. Rev. B}\ }\textbf {\bibinfo {volume}
  {83}},\ \bibinfo {pages} {045414} (\bibinfo {year} {2011})}\BibitemShut
  {NoStop}%
\bibitem [{\citenamefont {Williams}\ \emph {et~al.}(2011)\citenamefont
  {Williams}, \citenamefont {Low}, \citenamefont {Lundstrom},\ and\
  \citenamefont {Marcus}}]{Williams2011}%
  \BibitemOpen
  \bibfield  {author} {\bibinfo {author} {\bibfnamefont {J.~R.}\ \bibnamefont
  {Williams}}, \bibinfo {author} {\bibfnamefont {T.}~\bibnamefont {Low}},
  \bibinfo {author} {\bibfnamefont {M.~S.}\ \bibnamefont {Lundstrom}}, \ and\
  \bibinfo {author} {\bibfnamefont {C.~M.}\ \bibnamefont {Marcus}},\ }\href
  {\doibase 10.1038/nnano.2011.3} {\bibfield  {journal} {\bibinfo  {journal}
  {Nature nanotechnology}\ }\textbf {\bibinfo {volume} {6}},\ \bibinfo {pages}
  {222} (\bibinfo {year} {2011})}\BibitemShut {NoStop}%
\bibitem [{\citenamefont {Shylau}\ and\ \citenamefont
  {Jauho}(2014)}]{Shylau2014}%
  \BibitemOpen
  \bibfield  {author} {\bibinfo {author} {\bibfnamefont {A.~A.}\ \bibnamefont
  {Shylau}}\ and\ \bibinfo {author} {\bibfnamefont {A.-P.}\ \bibnamefont
  {Jauho}},\ }\href {\doibase 10.1103/PhysRevB.89.165421} {\bibfield  {journal}
  {\bibinfo  {journal} {Phys. Rev. B}\ }\textbf {\bibinfo {volume} {89}},\
  \bibinfo {pages} {165421} (\bibinfo {year} {2014})}\BibitemShut {NoStop}%
\bibitem [{\citenamefont {Wurm}\ \emph {et~al.}(2012)\citenamefont {Wurm},
  \citenamefont {Wimmer},\ and\ \citenamefont {Richter}}]{Wurm2012}%
  \BibitemOpen
  \bibfield  {author} {\bibinfo {author} {\bibfnamefont {J.}~\bibnamefont
  {Wurm}}, \bibinfo {author} {\bibfnamefont {M.}~\bibnamefont {Wimmer}}, \ and\
  \bibinfo {author} {\bibfnamefont {K.}~\bibnamefont {Richter}},\ }\href
  {\doibase 10.1103/PhysRevB.85.245418} {\bibfield  {journal} {\bibinfo
  {journal} {Physical Review B}\ }\textbf {\bibinfo {volume} {85}},\ \bibinfo
  {pages} {245418} (\bibinfo {year} {2012})}\BibitemShut {NoStop}%
\bibitem [{\citenamefont {Rycerz}(2012)}]{Rycerz2012}%
  \BibitemOpen
  \bibfield  {author} {\bibinfo {author} {\bibfnamefont {A.}~\bibnamefont
  {Rycerz}},\ }\href {\doibase 10.1103/PhysRevB.85.245424} {\bibfield
  {journal} {\bibinfo  {journal} {Physical Review B}\ }\textbf {\bibinfo
  {volume} {85}},\ \bibinfo {pages} {245424} (\bibinfo {year}
  {2012})}\BibitemShut {NoStop}%
\bibitem [{\citenamefont {Ouyang}\ \emph {et~al.}(2010)\citenamefont {Ouyang},
  \citenamefont {Yang}, \citenamefont {Xiao}, \citenamefont {Wu},\ and\
  \citenamefont {Xu}}]{Ouyang2010}%
  \BibitemOpen
  \bibfield  {author} {\bibinfo {author} {\bibfnamefont {F.}~\bibnamefont
  {Ouyang}}, \bibinfo {author} {\bibfnamefont {Z.}~\bibnamefont {Yang}},
  \bibinfo {author} {\bibfnamefont {J.}~\bibnamefont {Xiao}}, \bibinfo {author}
  {\bibfnamefont {D.}~\bibnamefont {Wu}}, \ and\ \bibinfo {author}
  {\bibfnamefont {H.}~\bibnamefont {Xu}},\ }\href {\doibase 10.1021/jp1028454}
  {\bibfield  {journal} {\bibinfo  {journal} {The Journal of Physical Chemistry
  C}\ }\textbf {\bibinfo {volume} {114}},\ \bibinfo {pages} {15578} (\bibinfo
  {year} {2010})}\BibitemShut {NoStop}%
\bibitem [{\citenamefont {Cagliani}\ \emph {et~al.}(2014)\citenamefont
  {Cagliani}, \citenamefont {Mackenzie}, \citenamefont {Tschammer},
  \citenamefont {Pizzocchero}, \citenamefont {Almdal},\ and\ \citenamefont
  {B{\o}ggild}}]{Cagliani2014}%
  \BibitemOpen
  \bibfield  {author} {\bibinfo {author} {\bibfnamefont {A.}~\bibnamefont
  {Cagliani}}, \bibinfo {author} {\bibfnamefont {D.~M.~A.}\ \bibnamefont
  {Mackenzie}}, \bibinfo {author} {\bibfnamefont {L.~K.}\ \bibnamefont
  {Tschammer}}, \bibinfo {author} {\bibfnamefont {F.}~\bibnamefont
  {Pizzocchero}}, \bibinfo {author} {\bibfnamefont {K.}~\bibnamefont {Almdal}},
  \ and\ \bibinfo {author} {\bibfnamefont {P.}~\bibnamefont {B{\o}ggild}},\
  }\href {\doibase 10.1007/s12274-014-0435-x} {\bibfield  {journal} {\bibinfo
  {journal} {Nano Research}\ }\textbf {\bibinfo {volume} {7}},\ \bibinfo
  {pages} {743} (\bibinfo {year} {2014})}\BibitemShut {NoStop}%
\end{thebibliography}

%

\end{document}